\documentclass[fleqn,usenatbib]{mnras}

\usepackage{newtxtext,newtxmath}

\usepackage[T1]{fontenc}

\DeclareRobustCommand{\VAN}[3]{#2}
\let\VANthebibliography\thebibliography
\def\thebibliography{\DeclareRobustCommand{\VAN}[3]{##3}\VANthebibliography}

\usepackage{graphicx}	
\usepackage{amsmath}	

\usepackage{amssymb}	
\usepackage{ulem}
\usepackage{comment}
\usepackage{color}
\usepackage[flushleft]{threeparttable}

\usepackage{tikz,xcolor,hyperref}

\definecolor{lime}{HTML}{A6CE39}
\DeclareRobustCommand{\orcidicon}{%
	\begin{tikzpicture}
	\draw[lime, fill=lime] (0,0) 
	circle [radius=0.16] 
	node[white] {{\fontfamily{qag}\selectfont \tiny ID}};
	\draw[white, fill=white] (-0.0625,0.095) 
	circle [radius=0.007];
	\end{tikzpicture}
	\hspace{-2mm}
}

\foreach \x in {A, ..., Z}{%
	\expandafter\xdef\csname orcid\x\endcsname{\noexpand\href{https://orcid.org/\csname orcidauthor\x\endcsname}{\noexpand\orcidicon}}
}

\definecolor{darkgreen}{rgb}{0.13, 0.55, 0.13}

\newcommand{\HII}{H{\sc ii}}

\newcommand{\kms}{km s$^{-1}$}

\title[Sub-kpc scale Gas Density Histogram]{Sub-kpc scale gas density histogram of the Galactic molecular gas: a new statistical method to characterise galactic-scale gas structures}
\author[R. Matsusaka et al.]{Ren Matsusaka\orcidA{},$^{1}$\thanks{E-mail:ren.matsusaka.jp@gmail.com}
Toshihiro Handa,$^{1,2}$\thanks{E-mail:handa@sci.kagoshima-u.ac.jp}
Yusuke Fujimoto\orcidB{},$^{3}$
Takeru Murase\orcidC{},$^{1,4}$
Yushi Hirata\orcidD{},$^{1}$
\newauthor 
Junya Nishi,$^{1}$
Takumi Ito,$^{5}$
Megumi Sasaki$^{1}$
and Tomoki Mizoguchi$^{1}$
\\
$^{1}$Department of Physics and Astronomy, Graduate School of Science and Engineering, Kagoshima University, 1-21-35 Korimoto,\\ Kagoshima, Kagoshima 890-0065, Japan\\
$^{2}$Amanogawa Galaxy Astronomy Research Center, Kagoshima University, 1-21-35 Korimoto, Kagoshima, Kagoshima 890-0065, Japan\\
$^{3}$Department of Computer Science and Engineering, University of Aizu, Tsuruga Ikki-machi, Aizu-Wakamatsu, Fukushima 965-8580, Japan\\
$^{4}$Faculty of Engineering, Gifu University, 1-1 Yanagido, Gifu 501-1193, Japan\\
$^{5}$Graduate School of Science and Technology, Kumamoto University, Kumamoto, 860-8555, Japan\\
}

\date{Accepted 2024 January 18. Received 2024 January 18; in original form 2023 August 07}

\pubyear{2023}

\begin{document}
\label{firstpage}
\pagerange{\pageref{firstpage}--\pageref{lastpage}}
\maketitle

\begin{abstract}
To understand physical properties of the interstellar medium (ISM) on various scales, we investigate it at parsec resolution on the kiloparsec scale. Here, we report on the sub-kpc scale Gas Density Histogram (GDH) of the Milky Way. The GDH is a density probability distribution function (PDF) of the gas volume density. Using this method, we are free from an identification of individual molecular clouds and their spatial structures. We use survey data of $^{12}$CO and $^{13}$CO ($J$=1--0) emission in the Galactic plane ($l = 10\degr$--$50\degr$) obtained as a part of the FOREST Unbiased Galactic plane Imaging survey with the Nobeyama 45m telescope (FUGIN). We make a GDH for every channel map of $2\degr \times 2\degr$ area including the blank sky component, and without setting cloud boundaries. This is a different approach from previous works for molecular clouds. The GDH fits well to a single or double log-normal distribution, which we name the low-density log-normal (L-LN) and high-density log-normal (H-LN) components, respectively. The multi-log-normal components suggest that the L-LN and H-LN components originate from two different stages of structure formation in the ISM. Moreover, we find that both the volume ratios of H-LN components to total ($f_{\mathrm{H}}$) and the width of the L-LN along the gas density axis ($\sigma_{\mathrm{L}}$) show coherent structure in the Galactic-plane longitude-velocity diagram. It is possible that these GDH parameters are related to strong galactic shocks and other weak shocks in the Milky Way.
\end{abstract}




\section{Introduction}
\label{sec:intro}
Star formation is a sequential process composed of many stages from diffuse gas to a protostar, and each process takes place in different spatial scales from a few kpc in a galactic disk to 0.01 pc in a protostar core. Thanks to sub-arcsec submillimetre observations, many investigations into the final process have been made in recent decades. On the contrary, the first stages of the interstellar medium (ISM) and molecular cloud complex have been less studied.

In these early stages, there should be some physical connection between the ISM properties in kpc and sub-pc scales because some quantitative relations, such as the Kennicutt--Schmidt (KS) law, are found over these scales \citep{Schmidt1959,Kennicutt1989}. Another suggestion is a close relationship between the galaxy's star formation rate, efficiency, gas distribution, kinematics, and energetics \citep{HandaETAL1990,GouliermisEtAl2010,MomoseETAL2010,GrashaETAL2017,FujimotoETAL2019,ZhangETAL2019,VillanuevaETAL2022,MaedaETAL2023,DemachiETAL2023}.

To address these early stages in star formation, statistics on the ISM density structure is a promising approach. The histogram of the gas density over an assigned area will give a big hint because it will provide information about the properties and evolution of the density fluctuation of the ISM, which is equal to the probability distribution function (PDF) of the gas density if the fluctuation is steady. Many observational investigations using a histogram of the column density, the so-called N-PDF, have been performed on the molecular cloud scale, or a few tens of pc \citep[e.g.][]{Stutz&Kainulainen2015,BurkhartETAL2015,LinETAL2017,PouteauETAL2023}. Analogous to N-PDF studies, some investigations have used the brightness distribution index \citep[BDI,][]{SawadaETAL2012a,SawadaETAL2012b,SawadaETAL2018}. The BDI is derived from the brightness distribution in a single velocity channel and calculated as the pixel fraction above an assigned threshold. However, although the BDI is the direct statistic index of observational data, it is too simplified to consider the density structure of the ISM. Therefore, it has been difficult to develop it to address the origin of the ISM density structure. Although with the N-PDF, it is necessary to convert the integrated intensity of a molecular line to the ISM density, it is a closer quantity to the simulation studies, as shown below.

Many of these studies have shown that the N-PDFs of a molecular cloud or a part of it have different shapes. For example, some N-PDFs show a log-normal (LN) component with a power-law tail (PL) in its high column density side \citep[e.g.][]{SchneiderETAL2013,SchneiderETAL2015,SchneiderETAL2015b,SchneiderETAL2016,SchneiderETAL2022,PokhrelETAL2016,MaETAL2020,WangETAL2020,JiaoETAL2022}. The PL is often observed at the star-forming site, which is interpreted as a result of gravitational contraction \citep[e.g.][]{KlessenETAL2000,Ballesteros-ParedesETAL2011,ChenETAL2018,KhullarETAL2021,BiegingETAL2022}. However, some recent investigations show that the N-PDF comprises not one but multiple LN components and less correlation with star-forming activities \citep{MuraseETAL2023}. The other approach, \citep[e.g.][]{LeroyETAL2017,LiuETAL2021}, modelled the molecular gas PDF in galaxies using large velocity gradient (LVG) radiative transfer calculations to estimate CO line ratios. They have established a framework linking global CO line ratios to the mean molecular hydrogen gas density and kinetic temperature. These studies highlight the importance of improved observational knowledge of the PDF shape.

The first observational studies on the ISM density structure in the galaxy were based on the near-IR extinction map \citep[e.g.][]{LombardETAL2008,KainulainenETAL2009}. However, this method can be affected by contamination from foreground and background emission in the galactic plane, leading to an artificially increased gas column density. This means that N-PDF studies derived from dust continuum observations are only effective for individual molecular clouds with no foreground or background emission. They may be less accurate in studying the ISM beyond the cloud scale. To address this regime, line emission data can separate the overlapping objects along the same line of sight.

Using the line data with distance estimation, we can assign the ISM location in a 3-dimensional voxel without contamination. The diffuse ISM under the accumulation process should be located in the outer envelope or the intercloud space. The diffuse component may be missed if we use the data after any cloud identification \citep[e.g. clump find, CPROPS, Dendrograms][]{WilliamsETAL1994,Rosolowsky&Leroy2006,RosolowskyETAL2008}. To include this component for investigation on the early stage of ISM accumulation processes, we obtained a statistic of the gas density without any cloud identification. We called it the Gas Density Histogram (GDH), a fractional volume of every density diagram without morphological structure identification. This approach is a great advantage for investigating the early stage of ISM accumulation processes. The density distribution or the histogram of the gas density is the same as the PDF if all the processes are steady. In this paper, we use the GDH as a term of the PDF because the ISM density structure should be evolving and may not be steady.

One of the pioneering works on the 3-dimensional GDHs of the diffuse ISM was conducted by \cite{Berkhuijsen&Fletcher2008}. They estimated the GDH of the atomic and ionised ISM toward pulsars and stars in the Milky Way. However, their sample was small, and their results indicated that the derived GDH was consistent with a single LN. Another pioneering work was carried out by \cite{HandaETAL2012} using CO data along the galactic equator, i.e., $b=0$. They found that the GDH shows a single LN without any PL. 

To reduce the Poisson noise for each density bin, more than 1000 pixels are required to make a GDH when the density structure is assumed to be uniform in an area. It means that an area for a GDH, i.e., the effective resolution of the GDH, is at least 30 times larger than the pixel resolution. Therefore, the GDH analysis requires data with large coverage and sharp pixel resolution. Our analysis contains sufficient data to cover more than 282 $\times$ 282 pixels by using the FUGIN dataset. Thanks to a sensitive non-biased survey with a large radio telescope, we can make GDHs for each section of the sky and investigate its variety, which may be connected to the galactic structure in the sub-kpc scale. In this paper, we present the GDH distribution over a large part of the galactic plane.

We organise this paper as follows. In Section \ref{sec:data}, we show the specification of the data. In Section \ref{sec:data_analysis}, we calculate the volume density using the FUGIN data. Section \ref{sec:results} shows our resultant GDHs in each area at an assigned velocity and the effects of beam resolution on the GDH shape. Section \ref{sec:discussion} discusses the differences in GDH shape at different places in the galactic plane using the multi-LN fitting to the resultant GDHs and the GDH parameter on the longitude--velocity diagram with our corresponding interpretations.

\section{data}
\label{sec:data}
\begin{figure*}
	\includegraphics[width=\linewidth,trim={10 190 25 -10}, clip]{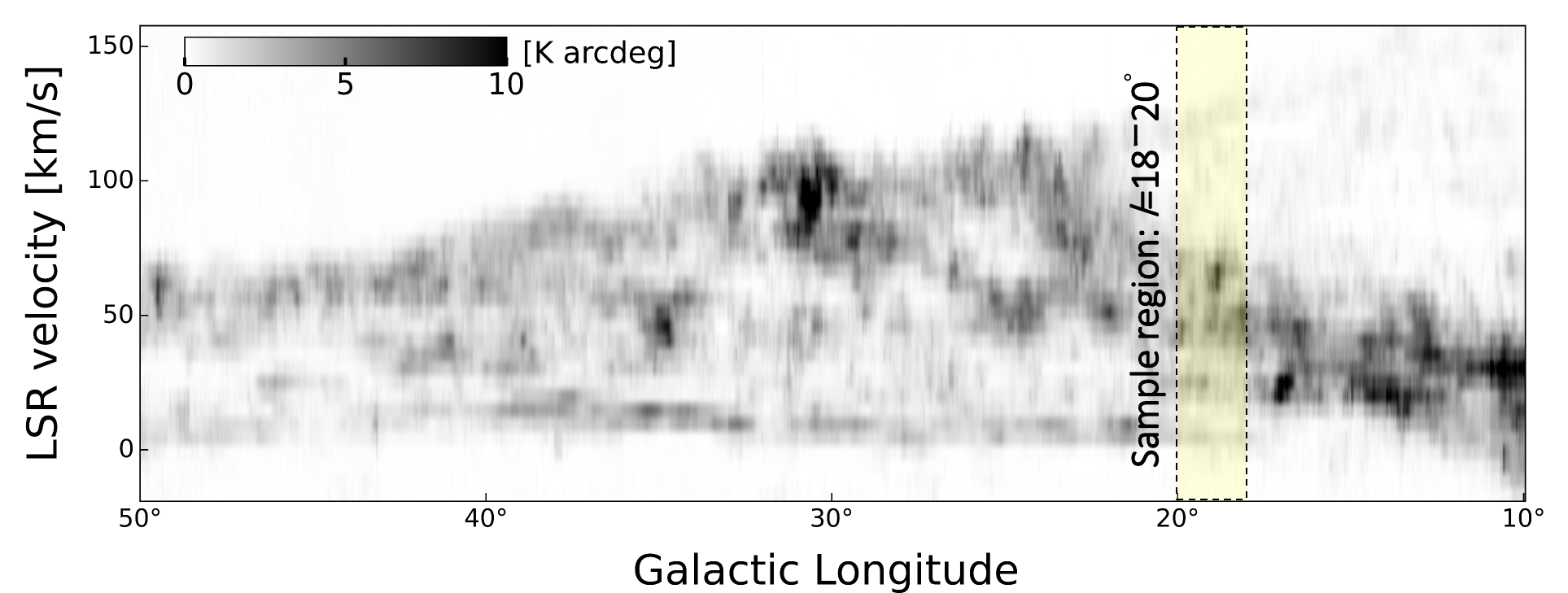}
    \caption{Longitude--velocity diagram in $^{12}$CO ($J$=1--0) data obtained from smoothed FUGIN datasets. The data were integrated with the galactic latitude over $|b| \leq 1.0\degr$.  The units on the intensity scale are K degrees. The vertical feature near $l = 31\degr$ is an artefact from the on-the-fly mapping. The yellow shaded region shows the $l = 18.0\degr$-- $20.0\degr$ region.}
    \label{fig:l-v}
\end{figure*}
We used $^{12}$CO and $^{13}$CO ($J$=1--0) datasets from the FUGIN project, which is an acronym of the FOREST Unbiased Galactic Plane imaging survey using the Nobeyama 45-m Telescope \citep{UmemotoETAL2017,ToriiETAL2019}. In this study, we used the data in the region of $10\degr \leq l \leq 50\degr$, $|b| \leq 1.0 \degr$
\footnote{FUGIN data is released by Japanese Virtual Observatory (JVO) \url{http://jvo.nao.ac.jp/}}. The angular and velocity resolutions of the published data are 20$\arcsec$ and 0.65 \kms, respectively. To improve the sensitivity, we smoothed over 8 spectral channels, with resulting velocity resolution of $\sim$ 5.2 \kms\, and realised two-dimensional spatial smoothing with a Gaussian function to obtain a resultant angular resolution of 1$\arcmin$. The smoothed resolution is sufficient for the density structure of the sub-kpc scale (see Section \ref{sec:resolution_effect}). Fig. \ref{fig:l-v} shows the longitude--velocity diagram in $^{12}$CO ($J$=1--0) data for the entire area covered in this study. The yellow-shaded region is representative of the whole area presented in Section \ref{sec:results}.

\section{Data analysis}
\label{sec:data_analysis}
In this paper, we present GDHs based on the volume density instead of the column density of the ISM. In this section, we show how to estimate the volume density and the volume of a voxel from the observed data.

\subsection{H$_{2}$ Column density}
The volume density is the column density divided by the length of the line-of-sight (LoS). As a first step, we estimate the column density of $^{13}$CO, $N(\mathrm{^{13}CO})$, from the $^{12}$CO ($J$=1--0) and $^{13}$CO ($J$=1--0) lines. We assume Local Thermodynamic Equilibrium (LTE) between these two transitions and the beam-filling factor ($\eta$) to be unity \citep{WilsonETAL2013book},
\begin{equation}
\centering
    N(\mathrm{^{13}CO}) = 3.0 \times 10^{14} \frac{T_\mathrm{ex}}{1 - \exp(-5.3 / T_\mathrm{ex})} \int \tau_{13}(v) dv, 
\end{equation}
where $dv$ is the velocity in \kms, $\tau_{13}$ is the $^{13}$CO opacity, and $T_{\mathrm{ex}}$ is the excitation temperature. Then, the $^{13}$CO optical depth $\tau$($^{13}$CO) at each voxel should be connected through the following equation:
\begin{equation}
\begin{split}
  \tau(^{13}\mathrm{CO}) = -\ln{ \left[ 1 - \frac{T_\mathrm{MB}(^{13}\mathrm{CO
  }) / 5.3 }{1 / \{{\exp(5.3 / T_\mathrm{ex})-1\}} - 0.16} \right] },\\T_\mathrm{MB}(^{13}\mathrm{CO
  })=\left(J_{13}(T_\mathrm{ex})-J_{13}(T_\mathrm{CMB})\right)(1-e^{-\tau_{13}}),
  \label{eq:TexFromTb13}
\end{split}
\end{equation}
where $T\mathrm{_{MB}}$($^{13}$CO) is the main beam temperature of $^{13}$CO ($J$=1--0) line, and $J_{13}(T)$ is the equivalent brightness temperature of $T$ in Rayleigh--Jeans approximation at the $^{13}$CO ($J$=1--0) frequency, 
\begin{equation}
\centering
    J_{13}(T) \equiv \frac{h\nu}{k}\left[\exp\left(\frac{h\nu}{kT}\right)-1 \right]^{-1} , 
\end{equation}
and $T_\mathrm{CMB} = 2.7$ K is the cosmic microwave background temperature.

The excitation temperature is primarily derived from the brightness temperature of $^{12}$CO, $T_\mathrm{MB}$($^{12}$CO) through the following equation, because of the large opacity
\begin{equation}
    T_\mathrm{MB}(^{12}\mathrm{CO})=J_{12}(T_\mathrm{ex})-J_{12}(T_\mathrm{CMB}),
\label{eq:TexFromTb12}
\end{equation}
where $J_{12}(T)$ is the equivalent brightness temperature of $T$ in Rayleigh–Jeans approximation at the $^{12}$CO ($J$=1--0) frequency. 

However, $T_\mathrm{ex}$ should be warmer than 15 K, which is estimated by many investigations of the temperature of molecular gas \citep[such as][]{Planck-CollaborationETAL2011,SokolovETAL2017,ShimoikuraETAL2019,MuraseETAL2022}. Therefore, we set $T_\mathrm{ex}=15$ K if $T_\mathrm{ex} < 15$ K from Eq. (\ref{eq:TexFromTb12}). We have confirmed that the results shown in the next section are not affected if we use different threshold temperatures.

We used the abundance ratio of H$_2$ to $^{13}$CO as [H$_{2}/^{13}$CO] = $3.8 \times 10^{5}$ \citep[e.g.][]{PinedaETAL2008}, although the different value gives a small shift for the resultant GDH along the density axis \citep[see also][]{MuraseETAL2023}. 

\subsection{Line-of-sight depth of a voxel and its volume density}
\label{kinematic-distance}
\begin{figure}
    \includegraphics[width=\linewidth,trim={80 160 150 165}, clip]{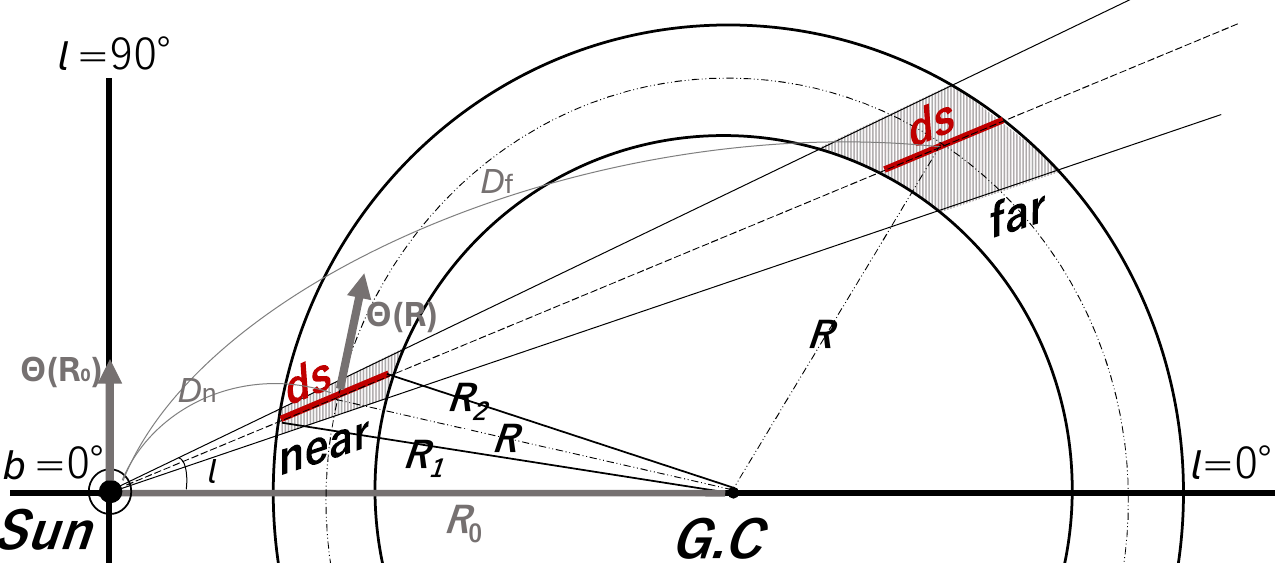}
    
    \caption{The rotation model provides exactly one distance $R$ from the galactic centre for every longitude $l$ and $V_{\rm{LSR}}$. There are two areas located at a galactocentric distance $R$, with longitude $l$ and latitude $b$ = 0$\degr$ from the solar system, as shown by the grey shades. Note that these cannot be separated from the $V_{\rm{LSR}}$.}
    \label{fig:kinematic-distance}
\end{figure}

For the line data, we use the kinetic distance to the ISM in the disk of the Milky Way \citep{Hulst&Oort1954}. To derive the kinetic distances, we assume the flat rotation curve and use the galactocentric distance to the Sun, $R_0$, and the rotation velocity of the local standard of the rest (LSR), $\it{\Theta_{\mathrm{0}}}$ as $R_0$ = 8.5 kpc and $\it{\Theta_{\mathrm{0}}}$ = 220.0 \kms\ \citep{Kerr&Lynden-Bell1986}.

The line of sight velocity against LSR ($V_{\rm{LSR}}$) of a molecular cloud (or interstellar gas) observed in the direction ($l$,$b$) rotating at the galactocentric distance $R$ is given by
\begin{equation}
    V_{\rm{LSR}} = \Big[\frac{R_0}{R}\it{\Theta(R)}-\it{\Theta_{\mathrm{0}}} \Big]\sin{l}\cos{b}.
\label{eq:kinematic-distance}
\end{equation}
We used $b$ = 0 because the observed range of the FUGIN dataset is only $|b| \le 1$. The heliocentric distance $D$ of the cloud can be given as the solution of 
\begin{equation}
    R^2 = D^2 + R_0^2 - 2DR_0 \cos{l},
\label{eq:galactcentric-distance}
\end{equation}
using $R$ given by Eq. \ref{eq:kinematic-distance}. An observed  $l-b-v$ voxel with a finite velocity width, $w$, has a finite range of the galactocentric distance between $R_1$ and $R_2$ in the given line of sight. At the given galactic longitude, $l$, the geometrical depth of a voxel, $ds$, at the near or far site is given by
\begin{equation}
    ds = \sqrt{R_2^2-R_0^2\sin^2{l}}-\sqrt{R_1^2-R_0^2\sin^2{l}},
\label{eq:depth}
\end{equation}
(see Fig. \ref{fig:kinematic-distance}). Using the depth $ds$ we can estimate the volume density of the gas from the column density;
\begin{equation}
    \rho(\rm{H_2}) = \frac{\it{N(\rm{H_2})}}{2\it{ds}}.
\label{eq:volume-density}
\end{equation}

We should note that each voxel is the sum of two separated volumes owing to the near/far ambiguity of the kinetic distance. Therefore, our data are the averages of two separate volumes.

The results are the almost same even when we use the values of $R_0$ = 7.92 kpc and $\it{\Theta_{\mathrm{0}}}$ = 238.9 \kms\ \citep[e.g.][]{VERACollaboration2020}. If we use a different rotation curve, the shape of the resultant GDH is not changed.

\section{results}
\label{sec:results}
This section presents the resultant GDH of each area along the galactic plane from FUGIN data, and the following sub sections present the artificial effects to the resultant GDH.

\subsection{The gas density histogram}
\label{subsec:gdh}
\begin{figure*}
    \includegraphics[width=\linewidth,trim={260 320 260 250}, clip]{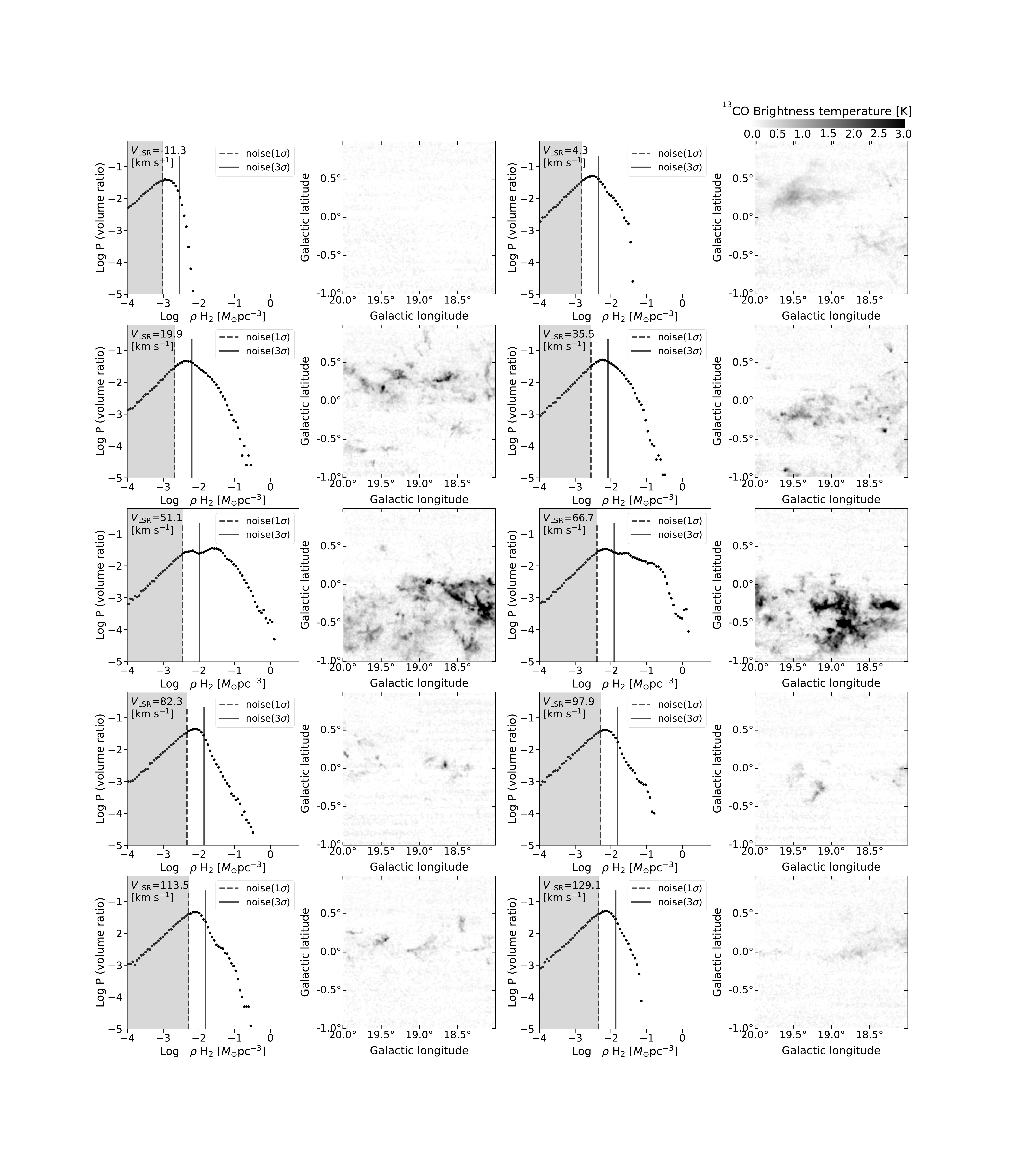}
    \caption{GDHs and intensity distribution on the sky of the corresponding areas, centred at $(l, b)=(19.0\degr, 0\degr)$ at different velocities. The total number of pixels is 282 $\times$ 282 (=79542). The two vertical dotted and dashed lines correspond to 1$\sigma$ and 3$\sigma$, respectively. Along the ordinate axis, each plot has an error due to the number of pixels for each log $\rho$ bin or the Poisson noise. However, it is not shown in this study as the effect is too small because the number of pixels in the GDH is sufficient. We show the line-of-sight velocity at the left-top corner of each panel. The full range of the grey scale is the same for all panels and is 0--3K.}
    \label {fig:rowGDH}
\end{figure*}

Previous studies on the N-PDF were performed for star-forming clouds and focused on its high-density end. However, the GDH is more powerful for diffuse and less dense gas because it does not require the identification of a ``cloud'' with a definite boundary. Therefore, we made the GDH of all ($l$-$b$-$v$) voxels in each geometrically square area. Fig. \ref{fig:rowGDH} shows 10 GDHs derived from channel maps of $18.0\degr < l < 20.0\degr, |b| \le 1\degr$ with $\Delta v=5.2$ \kms\ at different velocities as typical GDHs from the FUGIN data. The vertical lines are the sensitive limits estimated from 1$\sigma$ and 3$\sigma$ of the blank sky channel maps, respectively. Since the column density is primarily estimated from the $^{13}$CO intensity and the difference between the lengths of LoS is negligible in the assigned ($l$-$b$-$v$) area, the sensitivity limit for the volume density can be estimated from the noise level of $^{13}$CO. In Fig. \ref{fig:rowGDH}, we can easily find that GDHs in the density range over the sensitivity limit change from place to place. This suggests that the ISM density structure differs from place to place.

We note that the gas density in the GDH is $\it{not}$ a thermodynamical one because the volume filling factor should be much less than unity even for each sampling voxel. That is why we use $M_\odot$ pc$^{-3}$ instead of H$_2$ cm$^{-3}$ as the unit of the abscissa. The typical density is 10$^{-2}$ $M_\odot$ pc$^{-3}$, which is equal to $2.0\times 10^{-1}$ H$_2$ cm$^{-3}$, less than the CO critical density by three or four orders of magnitude \footnote{It should be noted that while the critical density of the CO ($J$=1--0) emission line is approximately 10$^{3}$ H$_2$ cm$^{3}$, it is optically thick, allowing it to be observed even at lower densities of approximately 10$^{2}$ H$_2$ cm$^{3}$.}. Our estimation indicates the density averaged over a several-pc scale voxel under the low-volume filling factor. A low filling factor is consistent with lower brightness temperatures in the $^{12}$CO ($J$=1-0) line than the typical temperature of molecular gas or $\sim$ 15 K.

Fig. \ref{fig:rowGDH} suggests that areas with different morphology of intense emission show a different GDH shape. The GDHs of areas with diffuse less-bright gas, such as those at $v=35.5$ and 113.5 \kms\, show a sharp peak just above the sensitivity limit and a steep decline at the dense side. In contrast, the GDHs of areas with compact, bright features, such as those at 51.5 and 66.7 \kms, have another peak or excess component on the dense side. The simplest GDH is that at $v=-11.3$ \kms; it shows a close shape of Gaussian noise on the blank sky. At $v=4.3$ and 129.1 \kms, the GDHs have real features over the Gaussian noise at the densest tail, and the channel maps there show a few real features. 

We found that each GDH shows a straight line below the sensitivity limit. This is due to the Gaussian noise on the blank sky, which is quantitatively consistent with the noise estimation from the data.

\subsection{Spatial resolution effect}
\label{sec:resolution_effect}
\begin{figure*}
	\includegraphics[width=\linewidth,trim={300 30 300 100}, clip]{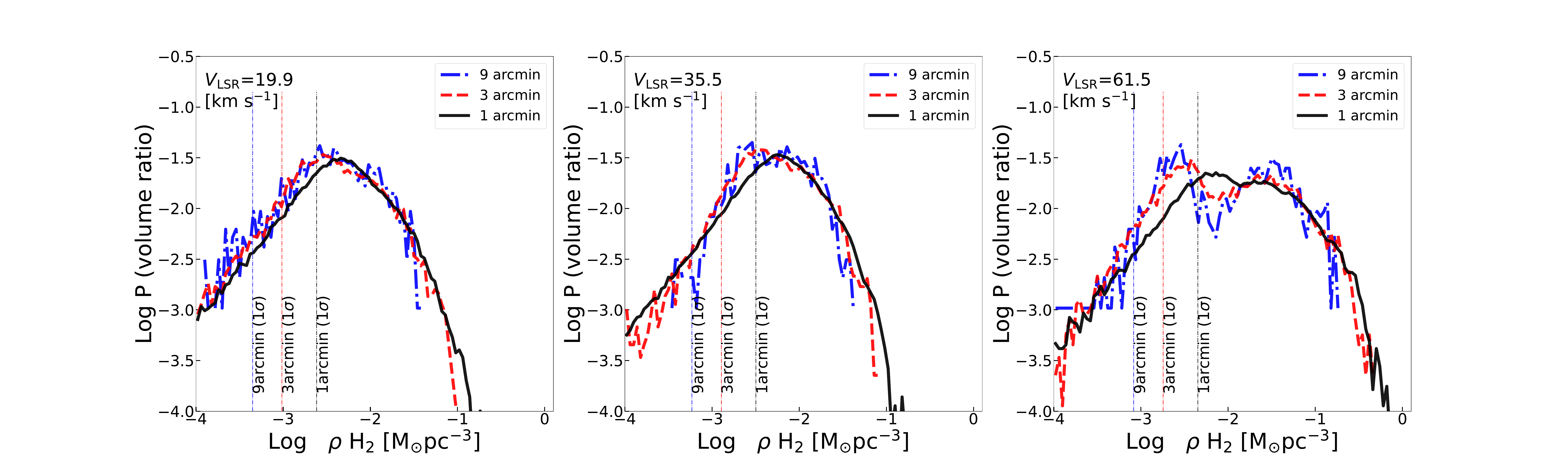}
    \caption{Resolution effect of the GDHs. The three GDHs in each panel are made from channel maps at 1$\arcmin$ resolution (black: solid line), 3$\arcmin$ resolution (red: dashed line), and 9$\arcmin$ resolution (blue: dash-dot line). These three panels correspond to different line-of-sight velocities of 19.9 \kms, 35.5 \kms and 61.5 \kms, respectively. The vertical dot-dashed line corresponds to 1$\sigma$.}
    \label{fig:RE}
\end{figure*}
GDHs are basically free from the effect of image resolution of the data because they are made without identification of the spatial structure of the ISM. However, they are actually affected because compact peaks must be diluted with an observation beam. We call it the resolution effect.

Surprisingly only a few investigations have been conducted on the resolution effect \citep[e.g.][]{SchneiderETAL2015b,SpilkerETAL2021}, and many previous investigations on N-PDFs do not consider it \citep[e.g.][]{SchullerETAL2017,ColomboETAL2022,MaETAL2020,MaETAL2022}. Therefore, we confirmed how seriously it affected our GDHs using smoothed images. The resolution effect is complex because it depends on the actual spatial structure of the ISM. We, therefore, made three GDHs using FUGIN $2\degr \times 2 \degr$ channel maps at $(l,b,v)=(19.0\degr, 0\degr, +19.9$ \kms,  +35.5 \kms, and +61.5 \kms) after smoothing to obtain $1^\prime, 3^\prime$, and $9^\prime$ angular resolution images (Fig. \ref{fig:RE}). 

The first significant difference is that the GDH made from the more smoothed image is the noisier. This is natural because of Poisson noise from the smaller number of voxels in the area. To obtain a reliable GDH, we need more than $10^3$ pixels in a single GDH area, as mentioned in Section \ref{sec:intro}. A much better voxel resolution is required to obtain a reasonable resolution of the GDH, or the size of GDH area, by a factor of 30.

The second significant difference is the highest end of the GDH. In almost all cases, the ISM has a spatial structure with compact clumps or cores. When these clumps and cores are smaller than the observed beam size, the resultant GDH misses the highest end, and their volumes contribute to lower-density bins. The GDH near the highest end is important for discussing the final stages of molecular cores, where the self-gravity of the gas is the major process. However, the resolution of our data of a few pc is too large to discuss such a process. We, therefore, do not focus on the high-density end of the GDH (see Section \ref{subsec:MultiLN}).

Besides these differences, the three GDHs are consistent, and the resolution effect is negligible between the sensitivity limit and $\log\rho($H$_2) \lesssim-0.8$. The resolution effect shown above may severely affect the GDH near the high-density end, and it has large Poisson noises.

The voxel resolution also changes the sensitivity limit because the ordinate values of our GDHs primarily come from the $^{13}$CO intensity. For the GDH shown in Fig. \ref{fig:noise}, the derived sensitivity limit from the rms noise level of the channel map is $\log \rho($H$_2) \simeq -2.6 M_\odot$ pc$^{-3}$, where the positive half of Gaussian noise with no real emission shows the peak in the $\log-\log$ plot. In particular, at 61.5 \kms, there are two significant peaks in the $9^\prime$ and $3^\prime$ images. To discuss whether this is real, we should consider the random noise due to the observation system. For more details, see Section \ref{subsec:substraction}.

Fortunately, most of our GDHs show a peak or plateau in the resolution effect-free range. We, therefore, can discuss the shape of GDHs with sufficient accuracy. In this study, we discuss the GDH from the $1^\prime$ data and its shape above the sensitivity and Poisson noise limits, which shows the real density structure.

\subsection{Velocity resolution effect}
\begin{figure*}

	\includegraphics[width=\linewidth,trim={300 30 300 100}, clip]{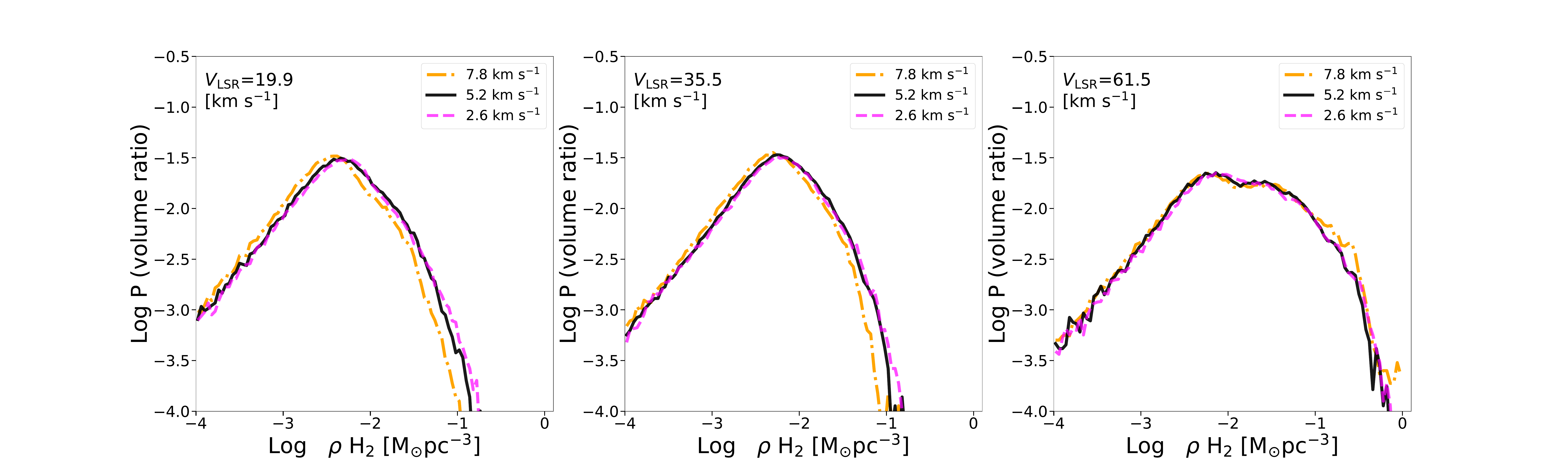}
    \caption{Velocity resolution effect of the GDHs. All GDHs in each panel are made from channel maps in 1$\arcmin$ angular resolution, but the velocity resolution is 2.6 \kms (magenta: dashed line), 5.2 \kms (black: solid line), and 7.8 \kms (orange: dash-dot line), respectively. The three panels correspond to the different line-of-sight velocities, as in Fig. \ref{fig:RE}.}
    \label{fig:vRE}
\end{figure*}
Although we convert the velocity difference to the distance, the cloud has internal turbulence. The kinematic distance is a nonlinear relation to the line-of-site velocity. Theoretical and observational studies suggest that velocity crowding affects the correlation analysis on the statistics of intensity fluctuations in Position--Position Velocity \citep[e.g.][]{Lazarian&Pogosyan2000,Lazarian&Pogosyan2004,YuenETAL2021}. We, therefore, evaluate the velocity resolution effect on the resultant GDH. 

We made three GDHs using FUGIN $2\degr \times 2 \degr$ channel maps at $(l,b,v)=(19.0\degr, 0.0\degr, +19.9$ \kms,  +35.5 \kms, and +61.5 \kms) after velocity smoothing of 2.6 \kms, 5.2 \kms, and 7.6 \kms (Fig. \ref{fig:vRE}). 

As shown in Fig. \ref{fig:vRE}, the global shape of each GDH is very similar. Therefore, there is no velocity crowding effect on the GDHs covered in this study. This looks strange because a molecular cloud has a line width of several \kms \citep[e.g.][]{SimonETAL2001,ColomboETAL2015,RosolowskyETAL2021,FujitaETAL2023}. However, it can be understood because the GDH resolution, i.e., the cell size of the GDH is $2\degr \times 2\degr$, or 300 pc $\times$ 300 pc at 8.5 kpc. This resolution is much larger than the typical cloud size, or 10 pc. Although each cloud has velocity width and random velocity against the pure galactic rotation, statistics of many clouds in the single GDH cell should be less affected by the velocity resolution. 

This also means that the GDH is less affected by the geometrical depth and velocity crowding. This is supported by the no systematic difference of GDHs near the terminal velocity (see Fig. \ref{fig:galactic_structure}).

\subsection{The effect of the beam filling factors and $^{13}$CO abundance}
\label{subsec:beam-filling-factor}
\begin{figure*}
	\includegraphics[width=\linewidth,trim={300 30 300 100}, clip]{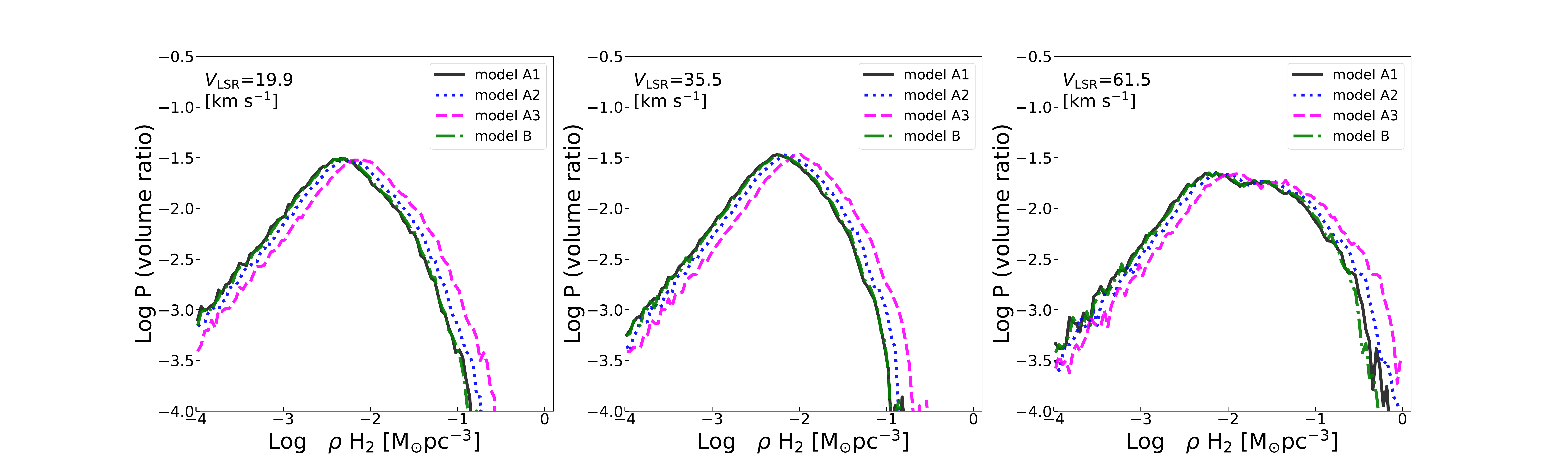}
    \caption{The four GDHs in each panel are derived using different assumptions of $T_{\rm{ex}}$ and beam filling factors. The angular resolution of all GDHs is the same as that of $1^\prime$. Model A1 is the solid (black, this work) line, Model A2 is the blue (dotted) line, Model A3 is the dashed (magenta) line, and Model B is the dash-dotted (green) line.}
    \label{fig:model}
\end{figure*}
\begin{table}

\centering 
    \caption{Beam-filling factor assumptions.}
    \label{table:model}
    \begin{threeparttable}
    \begin{tabular}{p{3em}|p{10em}|p{9em}}
    \hline
    \centering 
    Model&beam-filling factor&$T_{\rm{ex}}$\\ 
    \hline\hline
    \centering 
    A1 &  $\eta_{12}$ = 1.0, $\eta_{13}$ = 1.0  & optically thick $^{12}$CO\\
    \centering 
    A2 &  $\eta_{12}$ = 1.0, $\eta_{13}$ = 0.8 & optically thick $^{12}$CO \\
    \centering 
    A3 &  $\eta_{12}$ = 0.8, $\eta_{13}$ = 0.6  & optically thick $^{12}$CO\\
    \centering 
    B &  $\eta_{12}$ = not use, $\eta_{13}$ = 1.0  & 20 K (constant)\\
    \hline
    \end{tabular}
    \end{threeparttable}
\end{table}
As shown in Section \ref{sec:data_analysis}, we assume that the beam-filling factors of both CO lines are unity to derive our GDHs. In the case that the beam-filling factors are less than unity, the factors of $\eta_{13}$ and $\eta_{12}$ reduce both $T_\mathrm{MB}(^{13}\mathrm{CO})$ and $T_\mathrm{MB}(^{12}\mathrm{CO})$ from Eqs. \ref{eq:TexFromTb13} and \ref{eq:TexFromTb12}, respectively. However, the beam-filling factors have little effect on the GDH shapes, even if they are less than unity; they only cause some shift along the horizontal axis. 

This can be easily understood when the line intensities are much stronger than the CMB and $J(T) \simeq T$, i.e., $T_\mathrm{ex} \gg 5.5\ \mathrm{K}$. In this case, $T_\mathrm{ex}$ is proportional to $1/\eta_{12}$ and $\tau_{13}$ averaged over the beam increases proportionally to $\eta_{13}/\eta_{12}$. It means that $\tau$  for $^{13}$CO, estimated with $\eta_{12}=\eta_{13}=1$, is also applicable for other cases if $\eta_{12}=\eta_{13}$. Therefore, the resultant GDH is shifted along the horizontal axis and does not change its shape. To confirm the non-linear effects, we made additional GDHs using three models with different values of  $\eta_{12}$ and $\eta_{13}$ (Table \ref{table:model}). These GDHs are consistent with the expectation shown above (Fig. \ref{fig:model}). Model A and B are different sets of excitation temperature assumptions; A and B are the estimates $T_{\rm{ex}}$ from the optically thick $^{12}$CO and $T_{\rm{ex}}$ = 15 K constant, respectively. Models A1, A2, and A3 are the different beam-filling factors.

$^{13}$CO abundance has a similar effect on the GDH because it changes the conversion factor linearly from $N(^{13}\mathrm{CO})$ to $N(\mathrm{H}_2)$, which only produces a shift of the GDH along the horizontal axis in the log scale. The $^{13}$C abundance only shifts by a factor of 2 in the inner galaxy \citep[e.g.][]{MilamETAL2005,GiannettiETAL2014,JiaoETAL2021} without the galactic centre (FUGIN dataset). Even if the abundance is different for each GDH, their change is the shift of each GDH, but not a change in shape.

\section{Discussion}
\label{sec:discussion}

\subsection{Subtraction of random noise contribution}
\label{subsec:substraction}
\begin{figure*}
	\includegraphics[width=\linewidth,trim={300 30 300 100}, clip]{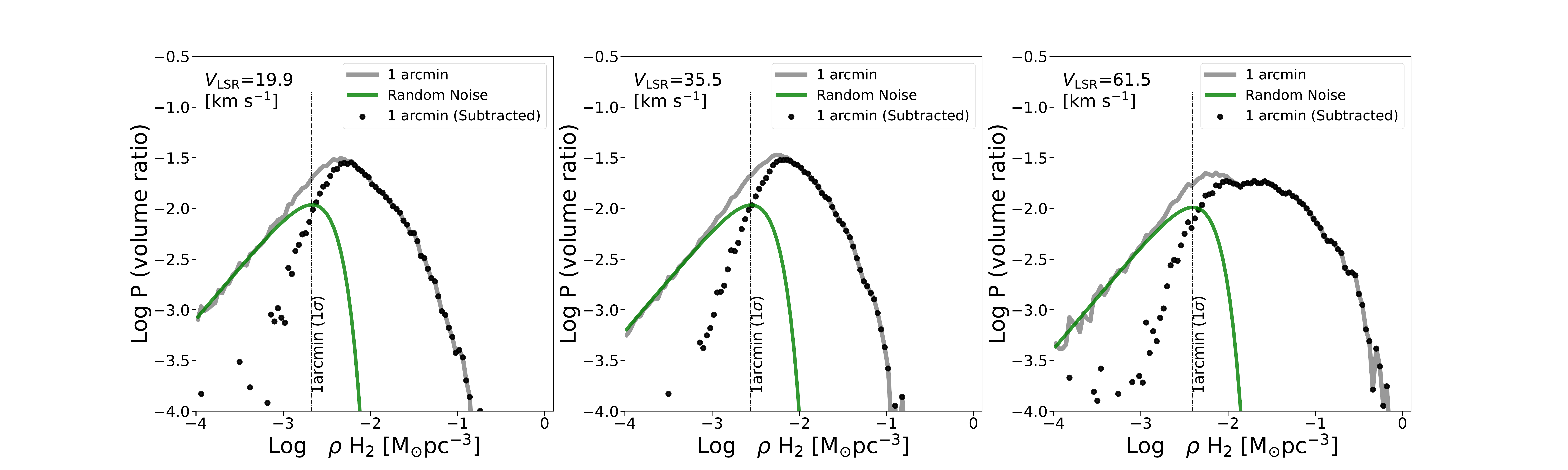}
    \caption{Same as Fig. \ref{fig:RE}, but after subtraction of the random noise contribution shown in thick green lines. The GDH before subtraction is shown in a solid grey line. The noise model is based on blank sky with the random noise estimated from the observations; the fitting parameter is only along the ordinate axis. The filled circle mark is subtracted from the random noise effects. The $^{13}$CO rms level is 0.07 K in $(l,b)$=(19.0\degr, 0\degr) areas.}
    \label {fig:noise}
\end{figure*}
The most striking feature of all GDHs is a straight line in the least dense region, although it is below the sensitivity limit. This comes, of course, from the random noise \citep[see also][]{OssenkopfETAL2016}, as mentioned in the previous section. In a log--log plot, Gaussian noise appears as a straight line with the power index of unity below the peak, which corresponds to its rms value. In addition, the effect of the noise decays rapidly beyond the peak (see Fig. \ref{fig:noise}). We can easily estimate its contribution to the observed GDHs with a single free parameter, i.e., the volume fraction of blank sky.

As the first step, we set a GDH component of blank sky with a Gaussian noise given by  
\begin{equation}
  P = {g(\rho)} = \frac{1}{\sqrt{2\pi}\sigma} \exp\left(-\frac{\rho^{2}}{2\sigma^{2}}\right),
\end{equation}
where $\sigma$ is the rms noise level. When we plot on the log--log plane, the horizontal axis should be described by $x=\log \frac{\rho}{\rho_\mathrm{norm}}$, i.e., $\rho=\rho_\mathrm{norm}\exp(x)$, where $\rho_\mathrm{norm}$ is the normalisation factor to set $x=0$. The vertical axis should be described by $y=\log g(\rho)$. Therefore, the curve of the Gaussian is described by
\begin{equation}
\label{noise}
  y = x - \log(\sqrt{2\pi}\sigma) - \frac{\rho_\mathrm{norm}^2 \exp(x)^2}{2\sigma^2},
\end{equation}
which is shown by a green line in each panel of Fig. \ref{fig:noise}.

We can fit by the noisy blank sky because it should occupy a large portion of the low-density region. Therefore, we can remove it from the observed GDH. However, we cannot apply it for the noise with very weak emission below the noise level, because we do not know the actual brightness distribution below the noise level. Beyond the noise level, the actual emission must be dominant. Therefore, we will discuss the profile beyond the noise level (Fig. \ref{fig:noise}: higher than the dashed line, Fig. \ref{fig:GDH}: non-shaded area).

Using this estimation, we subtract its contribution and determine whether there are components of GDH close to the sensitivity limit. It can give the ISM density structure in an outer region beyond the $\it{completeness\ limit}$ defined by previous N-PDF works on a molecular cloud \citep[e.g.][]{SchneiderETAL2015b,AlvesETAL2017,MaETAL2020,SpilkerETAL2021}. Although it is very important to investigate the density structure in the least dense envelope of ISM, it is beyond the scope of this paper.

\subsection{Multi log-normal fitting}
\label{subsec:MultiLN}
\begin{figure}
	\includegraphics[width=\linewidth,trim={10 40 10 45}]{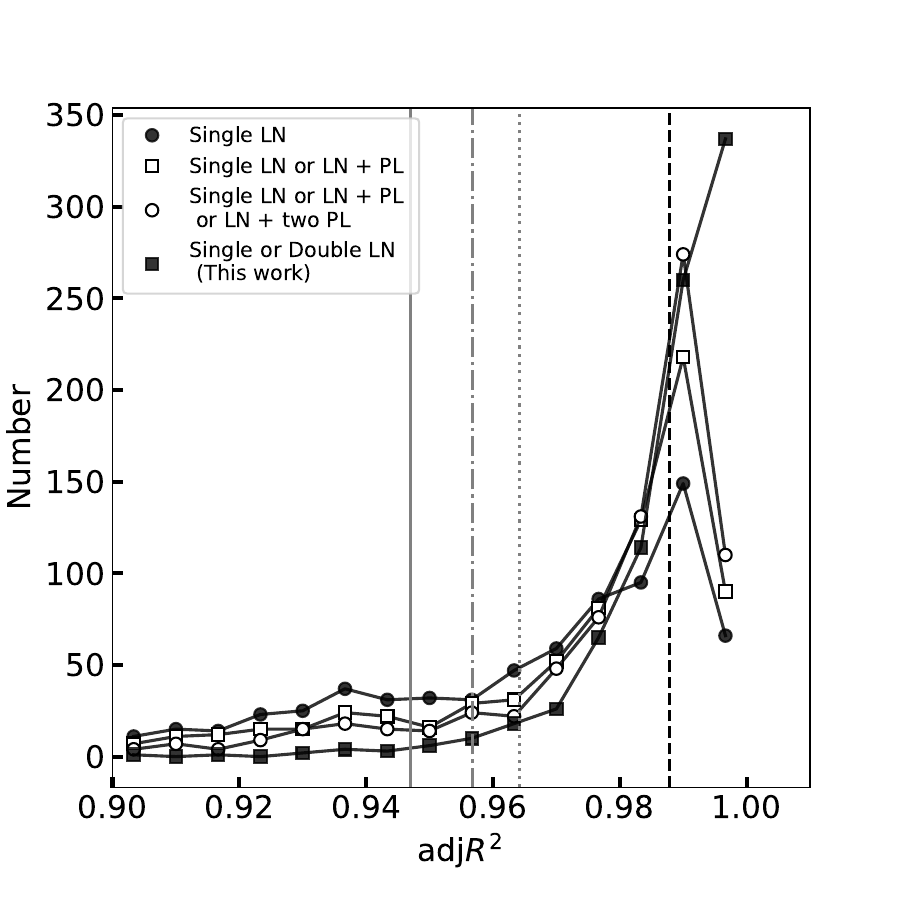}
    \caption{Results comparing adj$R^{2}$ with the four different models. The filled circle and solid line show results fitted with a single LN model, the open square and dashed-dotted line show those with a single LN or LN+PL model, and the open circle and dotted line show those with a single LN or LN+PL or LN+two PL model, and the filled square and dashed line represent those with single LN or double LN models (used in this paper). The vertical lines represent the mean value of adj$R^2$ over all areas with each of the models.}
    \label {fig:adjR2}
\end{figure}

\begin{figure*}
 \begin{minipage}{0.32\linewidth}
  \centering
  \includegraphics[width=\linewidth,trim={0 370 100 10}, clip, scale=0.333]
  {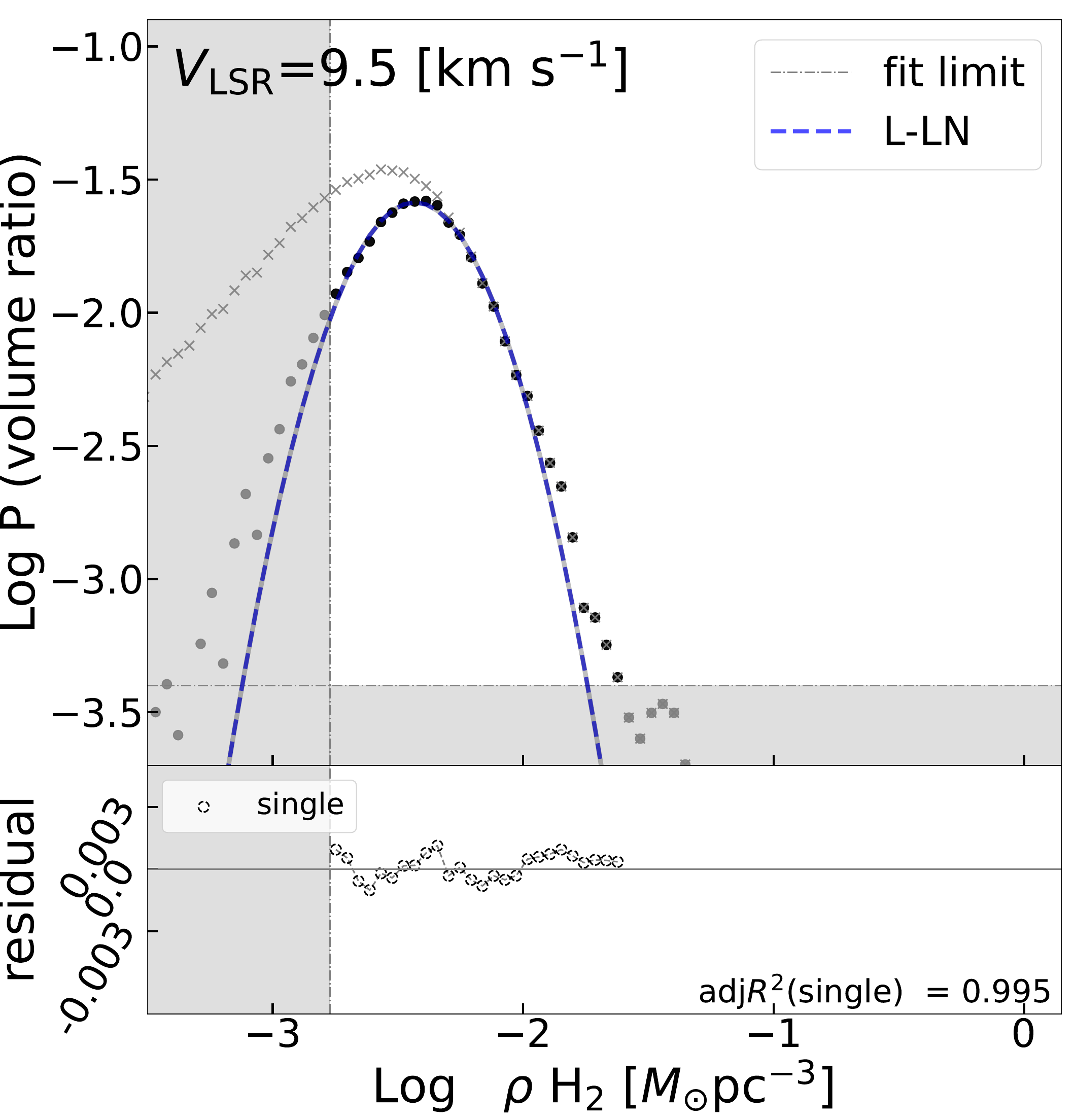}
 \end{minipage}
  \begin{minipage}{0.32\linewidth}
  \centering
  \includegraphics[width=\linewidth,trim={0 370 100 10}, clip, scale=0.333]
  {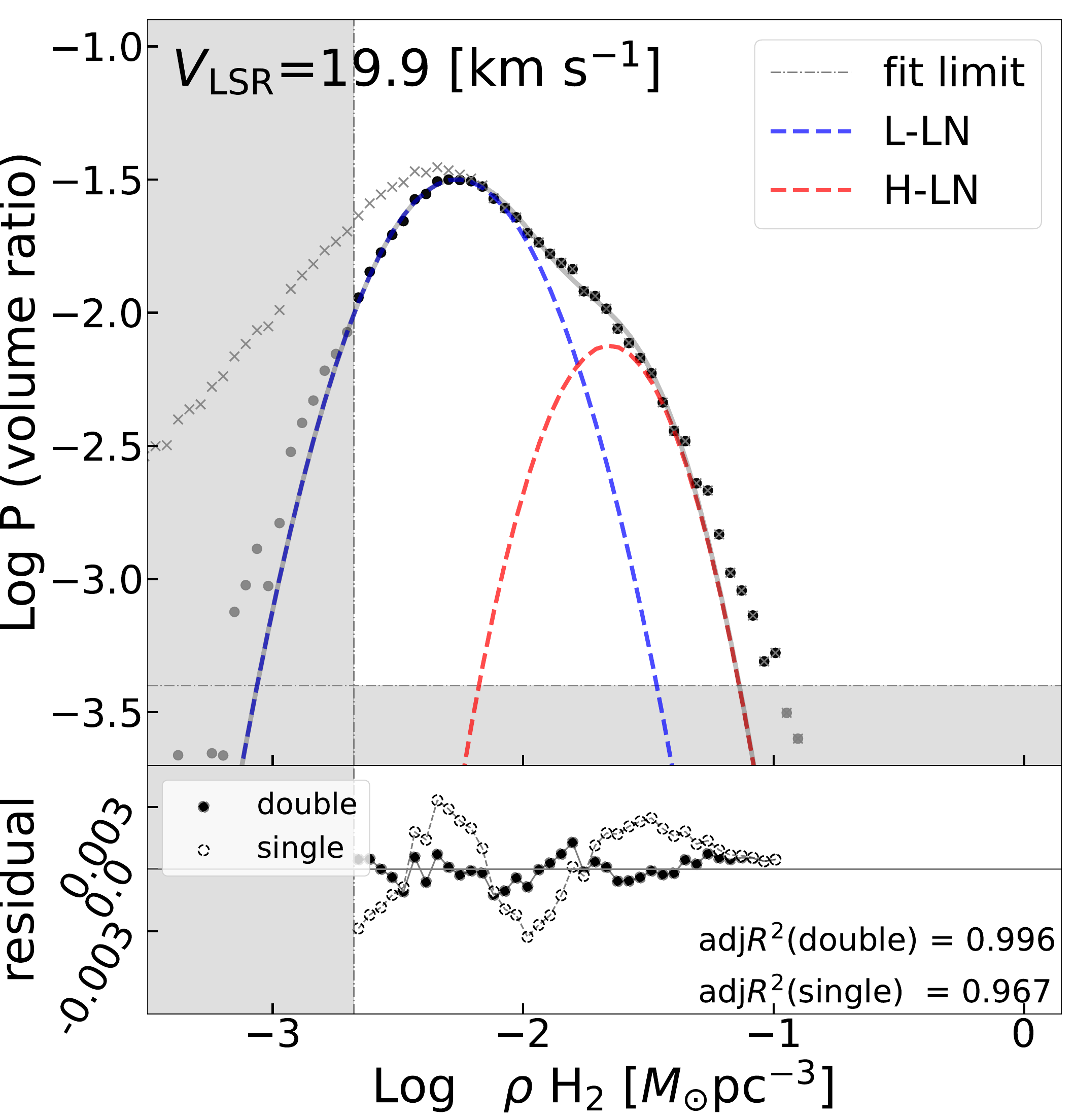}
 \end{minipage}
  \begin{minipage}{0.32\linewidth}
  \centering
  \includegraphics[width=\linewidth,trim={0 370 100 10}, clip, scale=0.333]
  {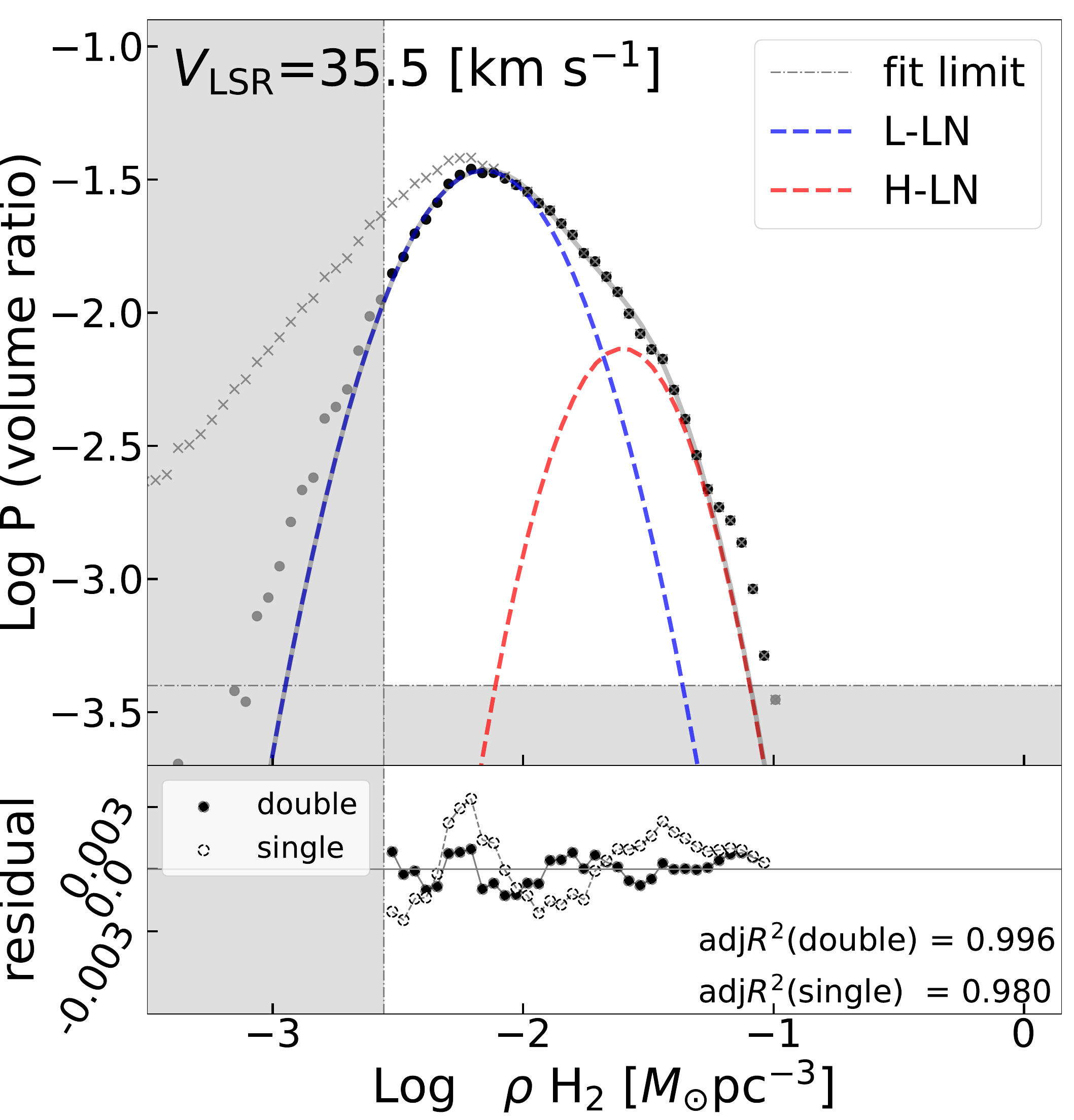}
 \end{minipage}
  \begin{minipage}{0.32\linewidth}
  \centering
  \includegraphics[width=\linewidth,trim={0 370 100 10}, clip, scale=0.333]
  {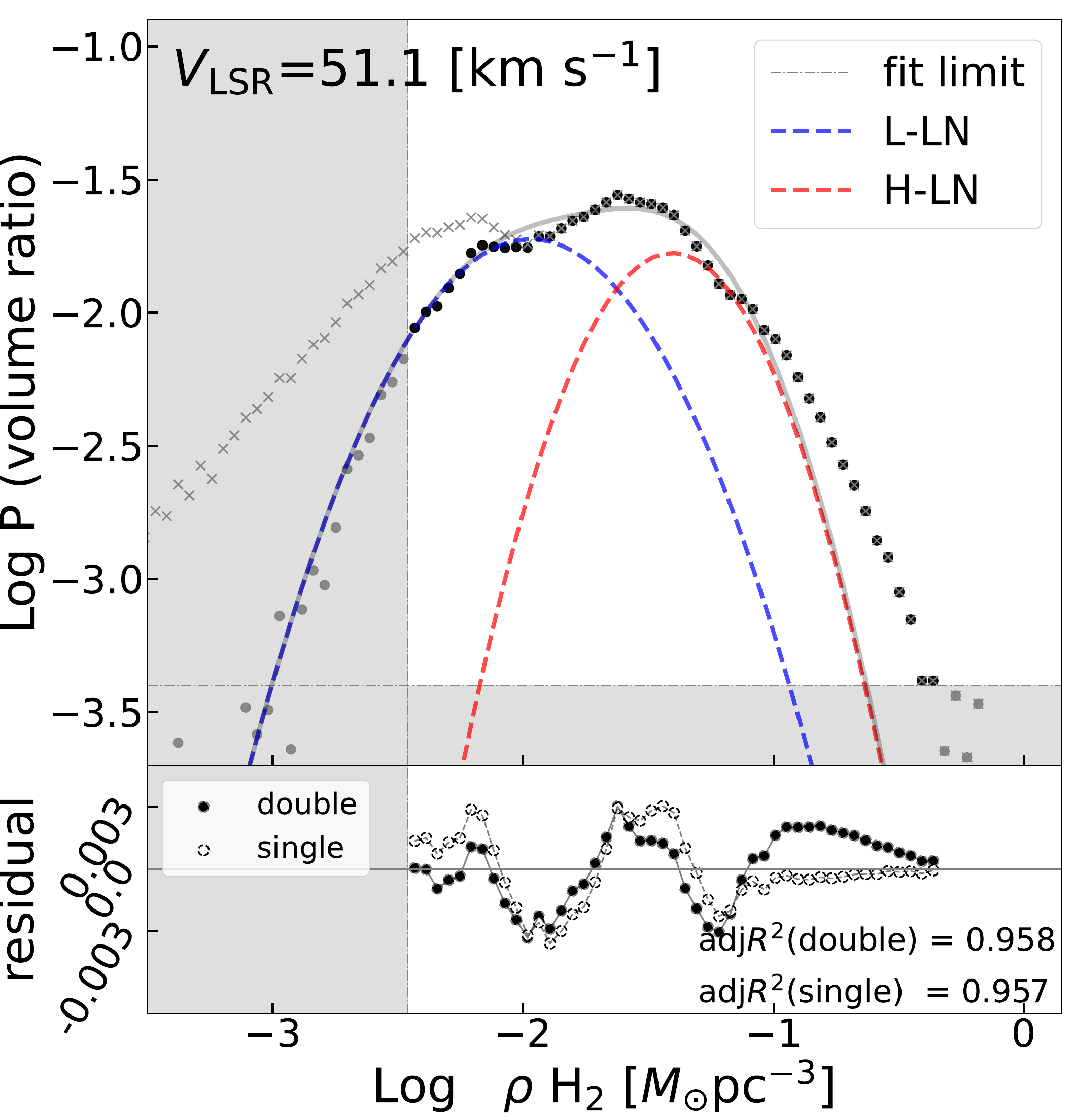}
 \end{minipage}
 \begin{minipage}{0.32\linewidth}
  \centering
  \includegraphics[width=\linewidth,trim={0 370 100 10}, clip, scale=0.333]
  {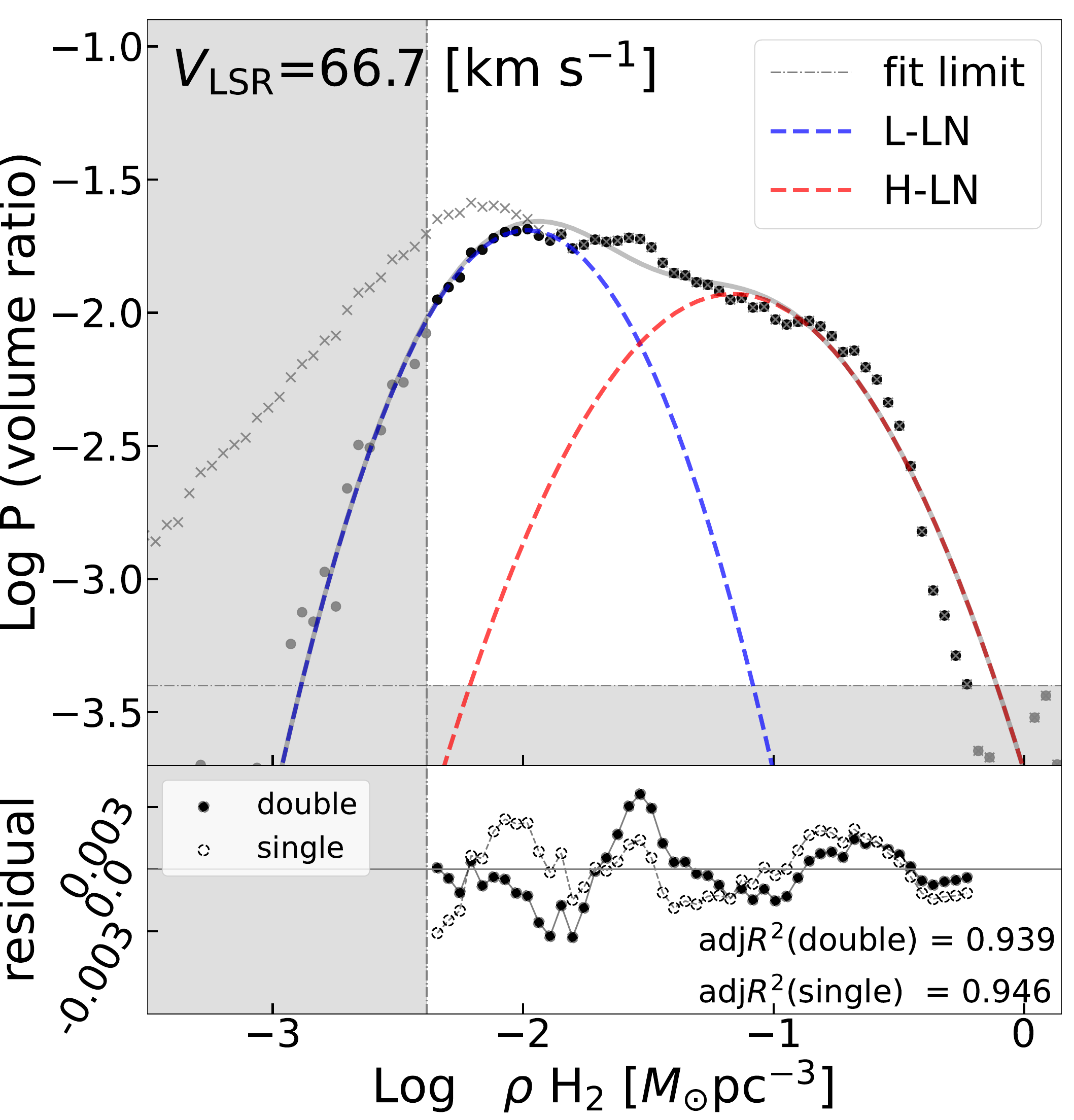}
 \end{minipage}
  \begin{minipage}{0.32\linewidth}
  \centering
  \includegraphics[width=\linewidth,trim={0 370 100 10}, clip, scale=0.333]
  {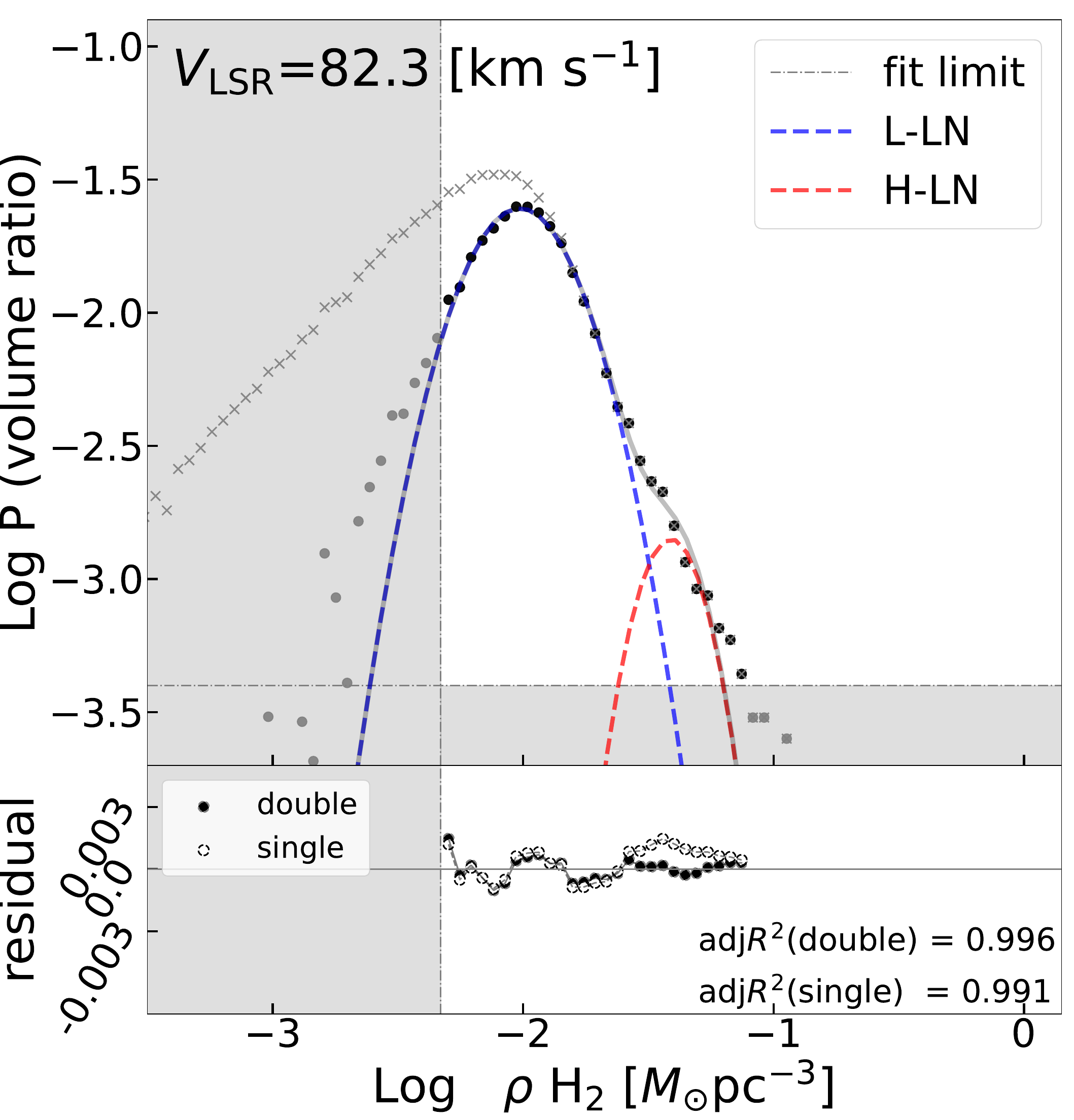}
 \end{minipage}
  \begin{minipage}{0.32\linewidth}
  \centering
  \includegraphics[width=\linewidth,trim={0 370 100 10}, clip, scale=0.333]
  {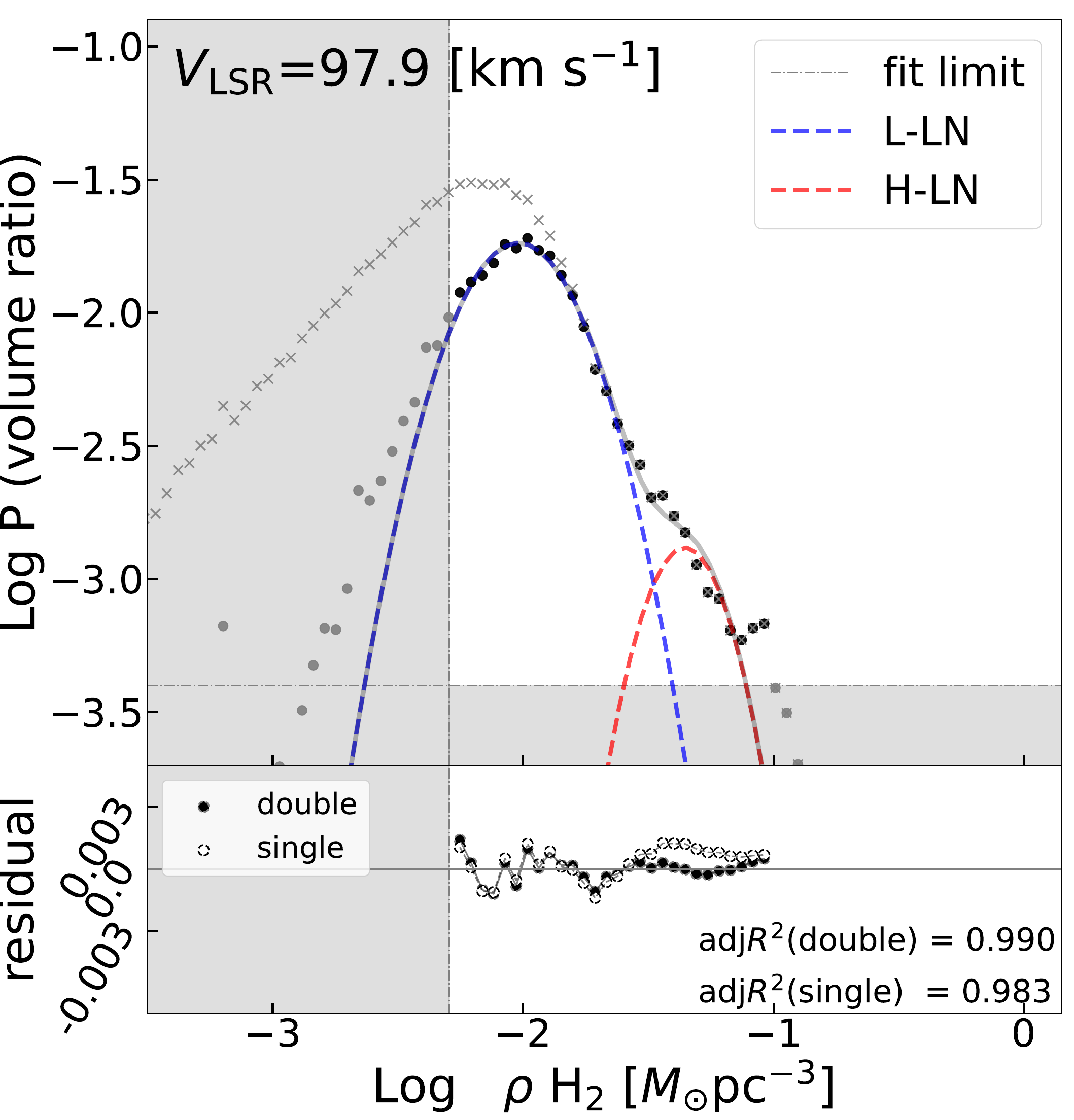}
 \end{minipage}
  \begin{minipage}{0.32\linewidth}
  \centering
  \includegraphics[width=\linewidth,trim={0 370 100 10}, clip, scale=0.333]
  {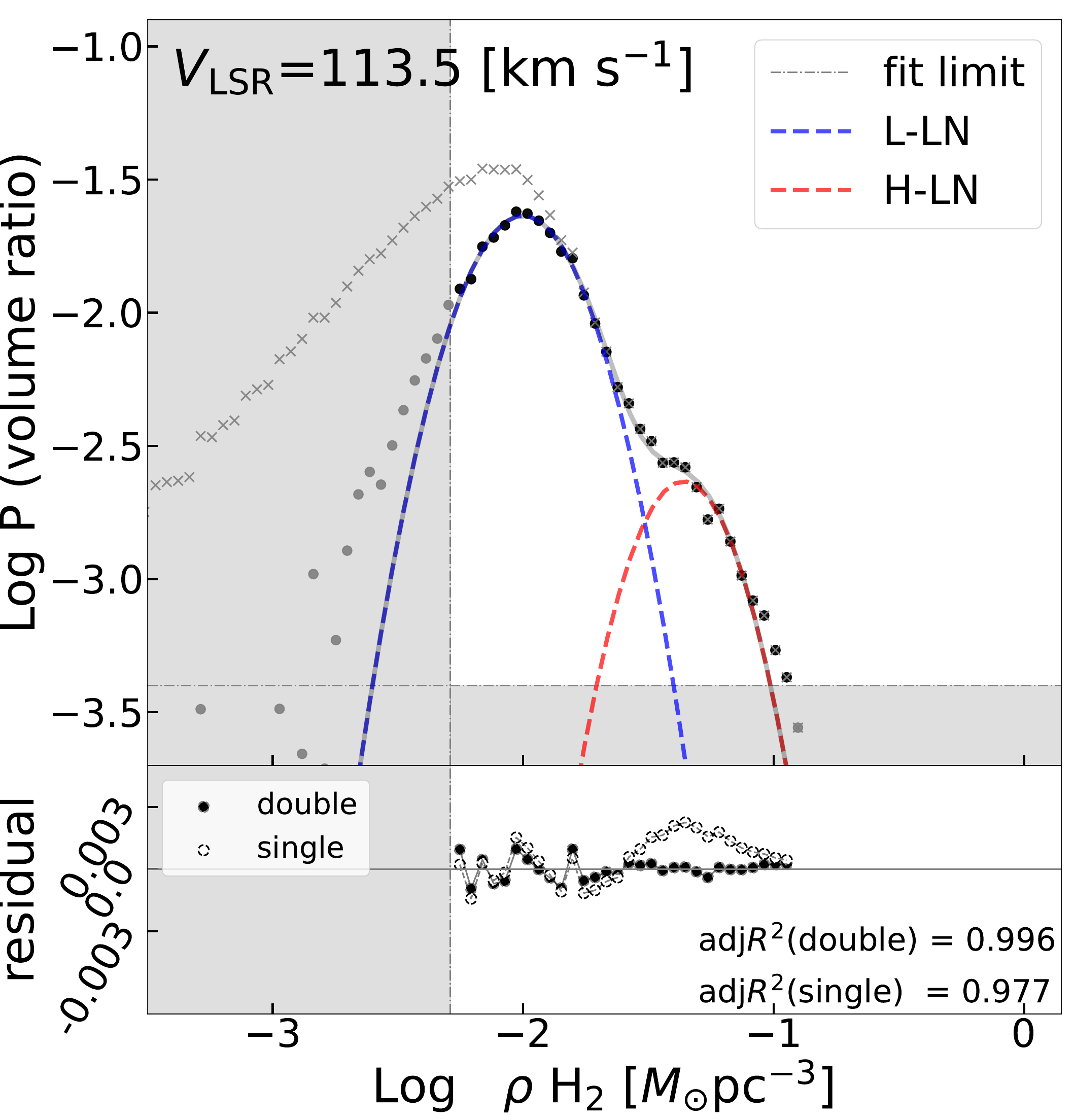}
 \end{minipage}
  \begin{minipage}{0.32\linewidth}
  \centering
  \includegraphics[width=\linewidth,trim={0 370 100 10}, clip, scale=0.333]
  {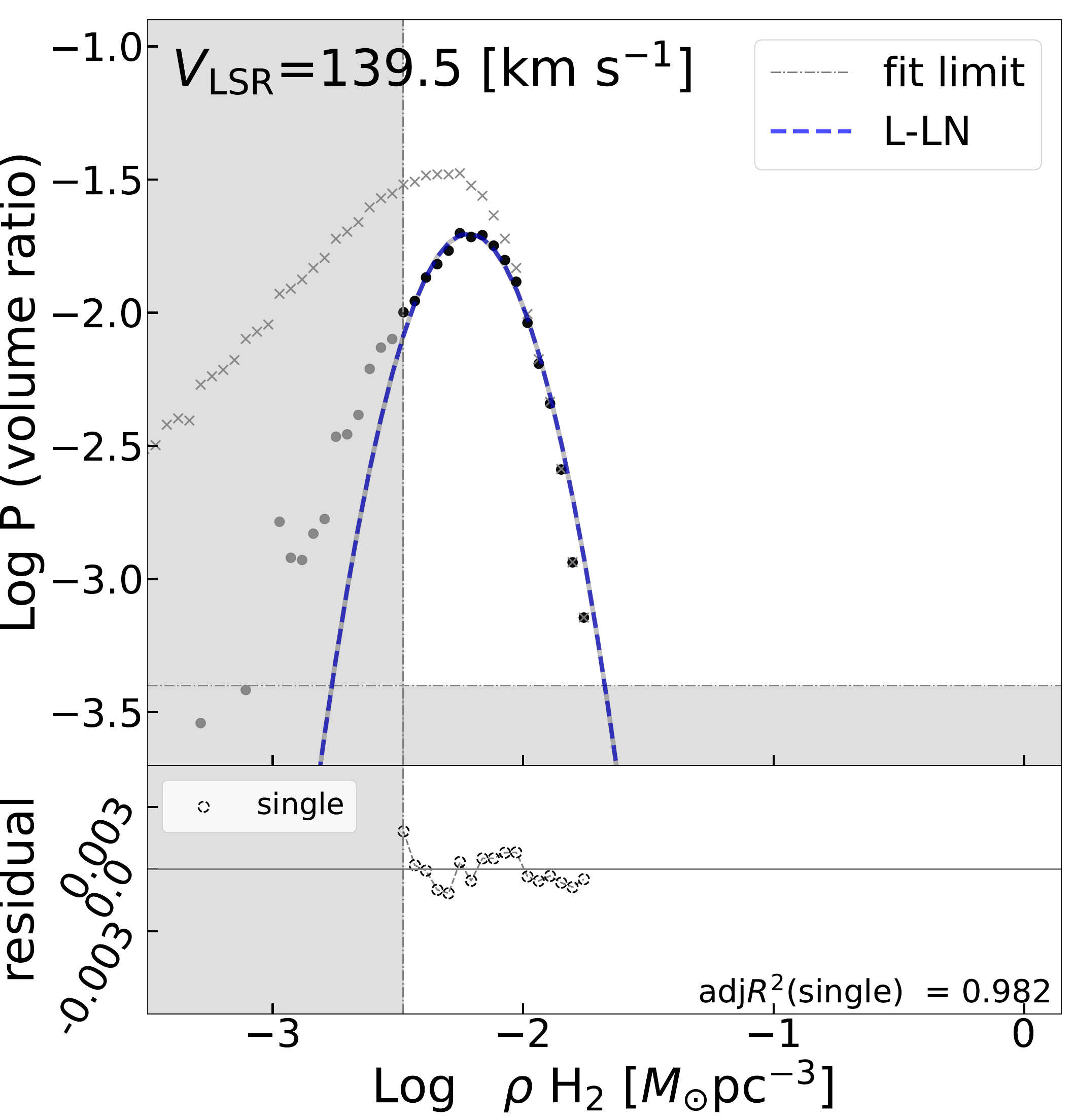}
 \end{minipage}
    \caption{GDHs after removing the Gaussian noise component and multi LN fitting at $l = 18.0\degr$--$20.0\degr, b = \pm 1\degr$. The vertical dashed lines correspond to 1$\sigma$. The L-LN and H-LN components are marked by blue and red dashed lines, respectively. Light grey dots show GDHs outside the discussion because of the low reliability of the noise removal procedure or poor pixels per bin. All GDHs were fitted using the least-squares method over 1$\sigma$ and sufficient pixel density per bin (N>10$^{1.5}$) for the random noise subtracted data. The $\log\rho-$P scale residuals of the single or double LN fitting are shown in an attached panel just beneath each GDH. In the residual plot, black circles show double LN residuals, and open circles show single LN residuals. In the lower right of the attached panel, we show adj$R^2$ for each of the models used.}
    \label {fig:GDH}
\end{figure*}
Although some GDHs can be fitted by a single LN\footnote{We fit the GDH in the $\log\rho-$P domain instead of the $\log\rho-\log$ P domain; the more weight is given for the larger P data.}, others require excess components on the single LNs (Fig. \ref{fig:GDH}). Most previous studies on molecular cloud's N-PDFs used one or two PL components to explain the excess. However, we propose that the second LN component naturally fit the excess because of the following two reasons. First, the double LN model expresses the observed GDH better than the LN plus PL (see Fig. \ref{fig:adjR2}). We evaluated the wellness of the fitting using the adjusted coefficient of determination (adj$R^2$) defined as follows:
\begin{equation}
    \label{adjR2}
     \mathrm{adj}R^{2} = 1-\frac{(n-1)\Sigma(y_{i,\mathrm{model}}-y_i)^2}{(n-p-1)\Sigma(y_i-\bar{y})^2},
\end{equation}
where $p$ is the number of free parameters, $n$ is the number of data points in the fit, y$_{i}$ is the volume fraction of the $i$th data point in the fit, y$_{i\mathrm{,model}}$ is the $i$th data point by each model fit, and $\bar{y}$ is the average of y. The adj$R^2$ closer to unity means that they can perfectly explain the observed GDH. This adj$R^2$ method is often used when the best number of parameters is unknown and it should be found by actual data \citep[e.g.][]{ShaoETAL2010,MannETl2014}. Many GDHs fitted by single or double LN models have values of adj$R^2$ above 0.95 and are much better than a single LN plus one or two PLs. We also found that the GDH has no systematic deviation from double LN fittings. 

Although a small number of GDHs can be fitted slightly better with the LN + PL model. We choose the double LN model for all GDH to simplify the interpretation, which will give a uniform model for the whole galaxy (Fig. \ref{fig:adjR2}).

Second, many previous works interpreted that these PL components are made by self-gravity. However, as \cite{MuraseETAL2023} pointed out, the self-gravity effect on GDH appears in 1 pc scale \citep{KhullarETAL2021}, which is much smaller than the pixel resolution in this work. 

We, therefore, fitted our GDHs with single or double LN distributions, following the approach of \cite{GazolETAL2013,Brunt2015,SchneiderETAL2022,MuraseETAL2023}. Therefore, the fitting function we used is the following:
\begin{multline}
    P(\rho) =  
    \frac{P_{\rm{L}}}{\sqrt{2 \pi}\sigma_{\rm{L}}} \exp \left(\frac{-\left(\log \rho-\log \rho_{\rm{L}}\right)^2}{2\sigma_{\rm{L}}^2} \right)  \\
    + \frac{P_{\rm{H}}}{\sqrt{2 \pi}\sigma_{\rm{H}}} \exp \left(\frac{-\left(\log \rho-\log \rho_{\rm{H}}\right)^2}{2\sigma_{\rm{H}}^2} \right),
\end{multline}
where $\rho_{\rm{L}}$ and $\rho_{\rm{H}}$ are the mean volume density of the high- and low-density LN distributions, respectively. We call these two components as L-LN and H-LN for the lower and higher density LN components, respectively. 

We also introduce a GDH parameter of the H-LN fraction. We calculated the volume ratios of the H-LN components to the total for each GDH defined by
\begin{equation}
    \label{eq:LH-LN_frac}
     f_\mathrm{H}=\frac{\int{P_\mathrm{H}(\rho)d\rho}}{\int{P_\mathrm{H}(\rho)d\rho}+\int{P_\mathrm{L}(\rho)d\rho}},
\end{equation}
which indicates the fraction of the H-LN component in the total gas. 

Some GDHs can be well-fitted without the second LN component. In this case, we call it a GDH with no H-LN component or $P_\mathrm{H}(\rho)=0$. 

We use an adj$R^2$ to perform model selection. More than 90$\%$ of all areas determined the number of LN components by this statistical method. However, the statistical method is not sufficient \citep[e.g.][]{MaETAL2022,SchneiderETAL2022}. We, therefore, also check the residuals of the best fit of the two models for each area. The residuals on the $\log\rho-$P scale are plotted below the GDHs, as shown in the panel attached to Fig. \ref{fig:GDH}. If the residuals are not distributed around zero, or if there is a large residual around the peak density that is the most highly weighted (see Fig. \ref{fig:GDH}, $v$ = 66.7 \kms), we consider the fitting to be not good. If the selected model based on adj$R^2$ unsatisfies the residual criterion, the GDH model is selected by eye. However, these situations occur only in a few percent of cases. 

Some GDHs throughout the galaxy suggest the presence of third LN-like components, for example, at $v$ = 51.1 \kms. However, these examples are limited, so we only consider up to the second components.

\begin{figure*}
\label{sec:l-v}
    \includegraphics[width=\linewidth,trim={175 37 175 70}, clip]{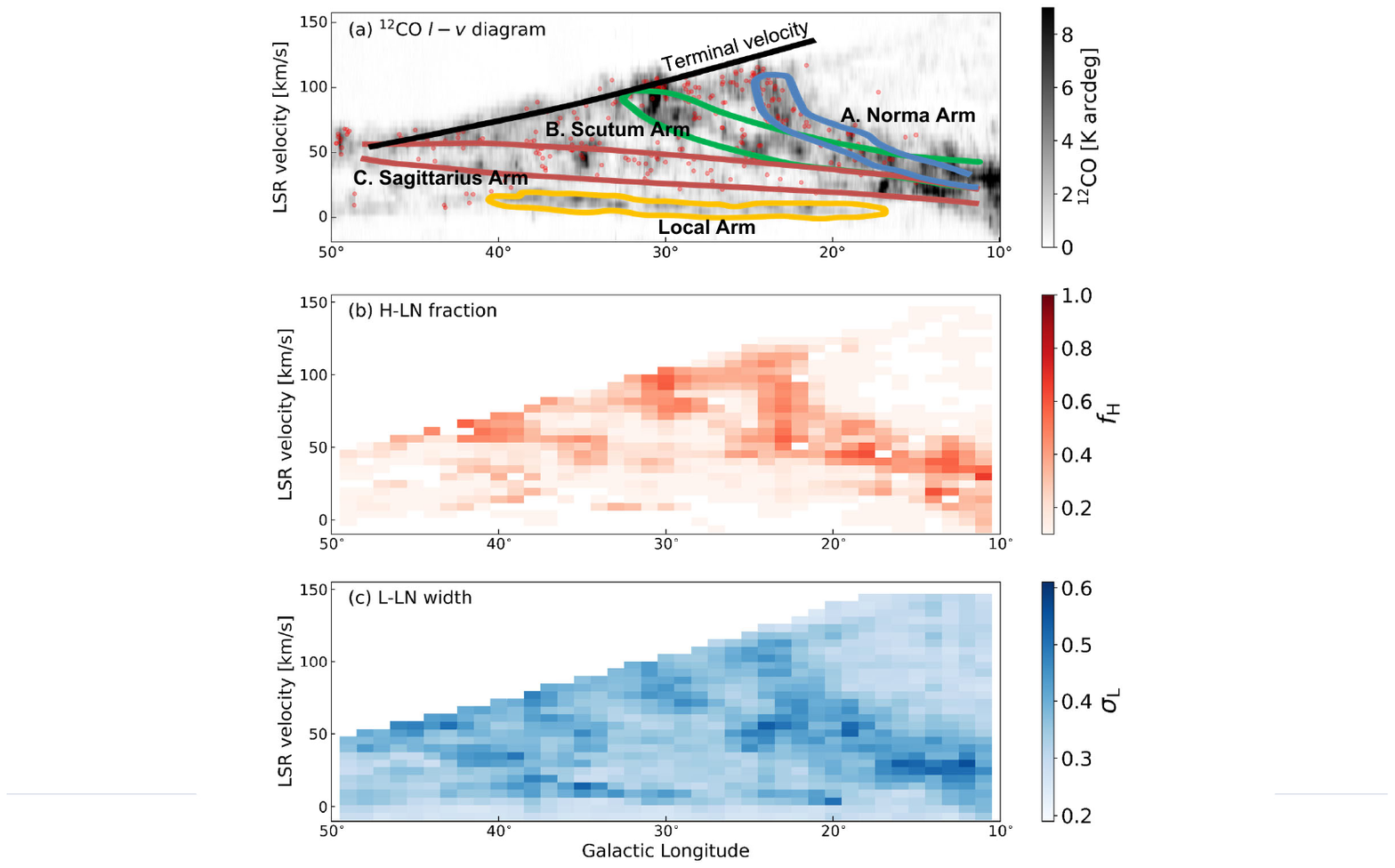}
    \caption{(a) $l-v$ diagram of the FUGIN $^{12}$CO $J$=1–0 data (same as Fig. \ref{fig:l-v}). The black line shows the curve of the terminal velocities. The coloured lines show the loci of the galactic arms constructed \citep{ReidETAL2016,ToriiETAL2019} and the \HII\ regions \citep[red points:][]{AndersonETAL2009}. (b) $l-v$ distribution of the $f_{\mathrm{H}}$ averaged over $\Delta l \times \Delta b$ = 2$\degr \times 2\degr$. The $2\degr \times 2\degr$ area where the H-LN cannot be defined (i.e., only diffuse gas, see Section \ref{sec:data_analysis}) is shown as $f_{\mathrm{H}}$=0. (c) The same as (b), but the colour is $\sigma_{\mathrm{L}}$.}
    \label{fig:galactic_structure}
\end{figure*}

\subsection{Interpretation of double LN components}
Using numerical simulation, \cite{Vazquez-Semadeni1994} showed that a GDH under a nearly pressureless supersonic flow without self-gravity is a single LN. They interpreted that the ``random walk’’ modulation of the gas density amplification makes Gaussian distribution along the logarithmic scale of the gas density, i.e., LN. In this case, the peak and the width of one LN component are the initial density of the gas and the evolutional stage of the random walk process or the modification factor at a single modification process, respectively.

We found that nearly all GDHs of $2\degr \times 2\degr$ areas can be expressed by double LNs (Fig. \ref{fig:GDH}). This means that the ISM in the single $2\degr \times 2\degr$ area is a mixture of two different density structures. However, we cannot separate them spatially. We can decompose the set of voxels into components based on the fitting and determine the fraction of each component at an assigned gas density. However, we cannot assign any voxel to a member of a given component. A method based on morphology with the help of the GDH analysis is required to separate them spatially in a sophisticated manner.

Although we cannot separate each component spatially, our results suggest that H-LN and L-LN correspond to compact and diffuse gas, respectively (a similar model is also suggested by \cite{GazolETAL2013}). We found that a single LN area shows only a few intense clumps embedded in the diffuse emission in its channel map, and a double LN area shows some intense clumps (see Fig. \ref{fig:rowGDH}).

The GDH comprises only LN, and no PL component is natural. It is not inconsistent with previous investigations, which showed the GDH or N-PDF of molecular clouds. We note that the resolution of our GDH is $2\degr$ or typically 300 pc, and any structure less than $20^{\prime\prime}$ or 0.8 pc can affect an observed GDH negligibly because this resolution is larger than the scale where gravitational collapse is the dominant process in the density evolution \citep[e.g.][]{ChenETAL2018,KhullarETAL2021}, as \cite{MuraseETAL2023} pointed out.

As we shown in Section \ref{sec:data_analysis}, our estimation of volume density is an average of two parts, i.e., near and far sides. However, this produces only small effects on the GDH because of the following two evidences. One is that we found no common feature in GDHs near the terminal velocity,  where the kinematic distance is given as the double root. The prominent features of the GDH parameters shown in Fig. \ref{fig:galactic_structure} are not limited to the $l-v$ area along the terminal velocity but continue to the other. The other evidence is that the GDHs are less affected by resolution effects both in velocity and space (see Section \ref{sec:results}). Therefore, the different source beam couplings do not severely affect the GDHs.

\subsection{Distribution of the GDH parameters on $l-v$ plane}
\label{subsec:gdh_lv_parameters}
GDHs can be expressed by some parameters because almost all observed GDHs can be well-fitted by single or double LN components, as shown in the previous section. We choose two independent parameters from five: the fraction of the H-LN component $f_\mathrm{H}$ and the half-width of the L-LN component $\sigma_\mathrm{L}$. Fig. \ref{fig:galactic_structure} shows their distribution and the $^{12}$CO integrated intensity on the $l-v$ plane. We found three features, named ridges A, B, and C, which connect from $(l, v)\simeq(10\degr$, 25 km s$^{-1})$ to $(25\degr, 100$ km s$^{-1})$, from $(l, v)\simeq(10\degr$, 25 km s$^{-1})$ to $(30\degr, 80$ km s$^{-1})$, and from $(10\degr, 10$ km s$^{-1})$ to $(50\degr, 60$ km s$^{-1})$, respectively. These ridges correspond to the spiral arms features traced in the CO intensity distribution (Fig. \ref{fig:galactic_structure}a). Ridges A, B, and C, are the Norma, Scutum, and Sagittarius arms, respectively \citep[e.g.][]{SandersETAL1985,DameETAL2001}. 
 
As it is well known, \HII\ regions are associated with spiral arms, and they are also associated with ridges A, B, and C \citep[e.g.][]{Georgelin&Georgelin1976,DownesETAL1980,AndersonETAL2012}. We note that there is no direct relation between $f_\mathrm{H}$ and $\sigma_{\rm{L}}$. Actually $f_\mathrm{H}$ and $\sigma_\mathrm{L}$ show the different distribution between 20$\degr$ and 40$\degr$ along $v \simeq 10$ km s$^{-1}$, which is the Local Arm. The Local Arm is a clear feature in the $^{12}$CO intensity from $(l,v) \approx (20.0\degr,10$ km s$^{-1}$) to $(40\degr, 10$ km s$^{-1})$ \citep[Fig. \ref{fig:galactic_structure}a; e.g.][]{ReidETAL2016,ReidETAL2019}. The Local Arm can also be traced in $\sigma_{\mathrm{L}}$, although no feature is found in $f_{\rm{H}}$. This suggests that the Local Arm is a physically different entity from the other arms. Although \cite{XuETAL2013} suggest that the Local Arm has similar kinematic properties as those found for the prominent spiral arms, it shows less active star formation \citep[see also][]{NakanishiETAL2016,SwiggumETAL2022}. This suggests that the broad $\sigma_\mathrm{L}$ is not made by star formation activity but by a kinematical effect in the sub-kpc scale. 

\subsection{Double LN components and the spiral arms}
\label{sec:what_ln}
\begin{figure}
    \includegraphics[width=\linewidth,trim={0 10 30 40}, clip]{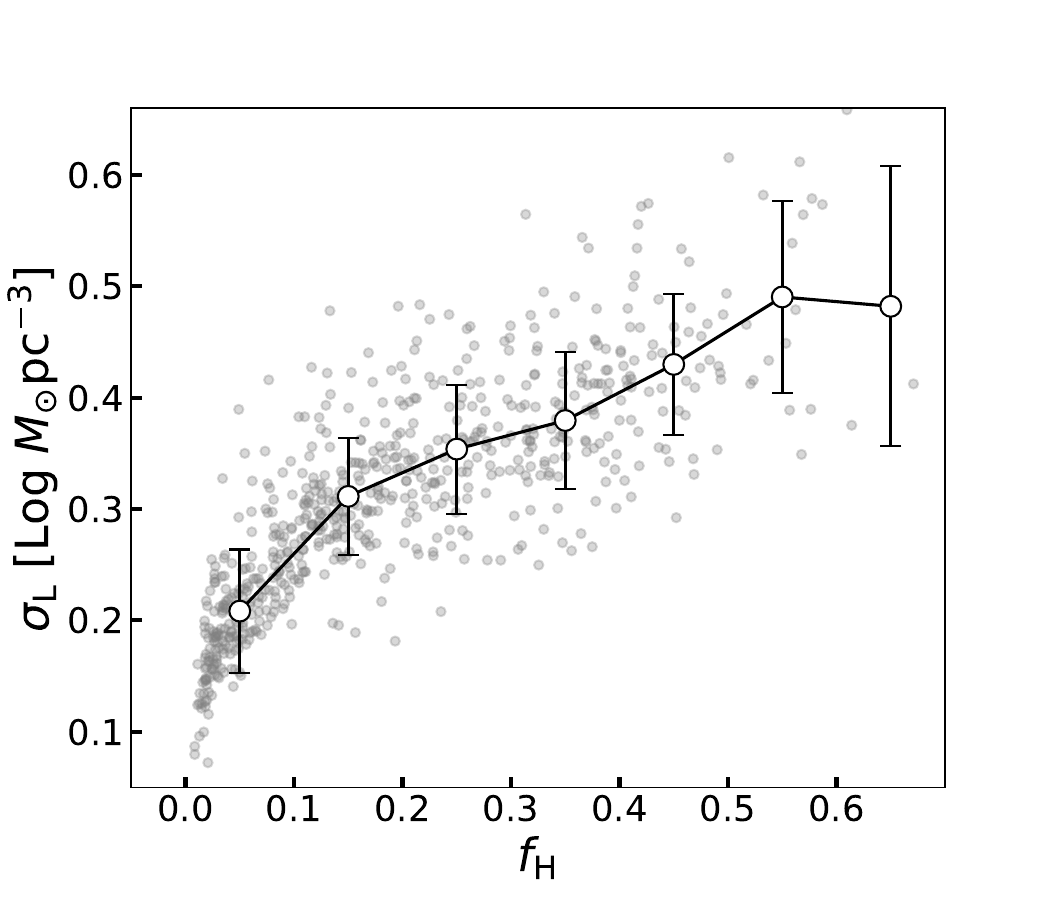}
    \caption{Correlation between $f_{\rm{H}}$ and $\sigma_\mathrm{L}$. The correlation coefficient is 0.80. The mean value and standard deviation of  $\sigma_\mathrm{L}$ in each 0.1 steps of $f_{\rm{H}}$ are over-plotted as white-open circles and bars, respectively. It is noted that the case $P_\mathrm{H}(\rho)=0$ is not plotted.}
    \label{fig:covariances}
\end{figure}
As shown in Section \ref{subsec:gdh_lv_parameters}, we found that the distribution of the CO line intensity is similar to those of the high $f_\mathrm{H}$ and wide $\sigma_\mathrm{L}$ (see Fig. \ref{fig:galactic_structure}b,c). What do they suggest? 

Based on the interpretation by \cite{Vazquez-Semadeni1994}, an LN component is made by a ``random walk’’ process in the ISM. A typical ``random walk’’ process consists of multiple shock passages produced by random supersonic turbulence. This shock must be on a smaller scale than the size of a GDH area. This is the natural interpretation of a single LN area. However, many areas show double LNs. Double LNs cannot be produced by the same mechanism as that of a single LN. Another process to make additional LNs with different peak densities is to give a different initial or mean density in the same area. What is this process?

The gravitational collapse is inadequate. Numerical simulation suggests it happens on much smaller scale \citep[e.g.][]{ChenETAL2018,KhullarETAL2021}. Although star formation feedback may compress the ISM, it is also inadequate. The feedback can affect the surrounding ISM only in a 1 pc order \citep[][and references therein]{RumbleETAL2021,MuraseETAL2022}. Furthermore, \cite{EgusaETAL2018} suggest that there is no significant effect of supernovae (SNe) on the kpc scale GDH. However, more detailed studies of nearby galaxies using pc scale resolution observations and simulations are needed to discuss the effect of the SNe.

A shock on a similar or larger scale than a GDH area is promising. The passage of the galactic shock \citep{Fujimoto1966,Fujimoto1968,WilliamsETAL1994} is a single shot and can produce coherent compression, which produces a shift of an LN component. Therefore, it can produce H-LN from L-LN efficiently. In this case, high $f_\mathrm{H}$ at the major spiral arm can be interpreted naturally \citep[e.g.][]{Wilson&Scoville1991,TosakiETAL2007}.

A wider $\sigma_\mathrm{L}$ is produced by more steps of density modification or by a larger change at each modification step. We cannot distinguish them only from the GDH shape. In the former case, $\sigma_\mathrm{L}$ shows the evolutional stage of L-LN \citep{WardETAL2014}. The width should be wider along the gas flow. Unfortunately, we cannot see it in galactic data due to the near--far ambiguity. GDH analysis of nearby galaxies using sensitive and high spatial resolution data will provide the answer.

In the latter case, the density modification is activated on a comparable or larger scale than a GDH area. The galactic shock may activate the small-scale shock via the cascading process. However, this mechanism must work in a short time scale because a wide $\sigma_\mathrm{L}$ has no systematic shift from the spiral arms traced as intensity ridges, which should be associated with the galactic shock. To clarify this, time evolution studies using large-scale and high-resolution ISM simulations with small CO clumps are required.

Fig. \ref{fig:covariances} shows a relation between $f_\mathrm{H}$ and $\sigma_\mathrm{L}$. There are correlations but large variance between $f_\mathrm{H}$ and $\sigma_\mathrm{L}$. This implies that the point-to-point correlations are poor. However, as can be observed in Fig. \ref{fig:galactic_structure}, both $f_\mathrm{H}$ and $\sigma_\mathrm{L}$ are well associated with the spiral arms on the $l-v$ plane. We guess this is due to a slightly different way of the association. As one important result, 

we found that the Local Arm is different from the major spiral arms. It shows wide $\sigma_\mathrm{L}$ but low $f_\mathrm{H}$. On the contrary, the part of $30\degr \lesssim l \lesssim 35\degr$ of ridge C shows high $f_\mathrm{H}$ but narrow $\sigma_\mathrm{L}$. They show no direct relation between high $f_\mathrm{H}$ and wide $\sigma_\mathrm{L}$. This gap is beyond our study. To clarify it, we need a more detailed investigation and it should be difficult to address this issue only with the Milky Way data. We plan to address this issue using a nearby face-on galaxy, and it will be presented in the forthcoming papers.

\section{Conclusions}
\label{sec:cunclusion}
We used all of the inner galaxy side data from FUGIN to investigate the characteristics of the gas density distribution in the Milky Way. Our conclusions are summarised as follows:

\begin{enumerate}
    \item We present the volume density histograms (GDHs) of $2\degr \times 2\degr$ areas of the inner disk of the Milky Way Galaxy using FUGIN data with the kinematic distance.
    
    \item All GDHs above the sensitivity limit can be well-fitted with single or double LN distributions. 
    \item The two GDH parameters, $f_{\mathrm{H}}$ and $\sigma_\mathrm{L}$, show three or four coherent enhancements on the $l-v$ plane. The pattern well traces the prominent spiral arms. However, the local spiral arm is traced by the wide $\sigma_\mathrm{L}$ but not by the high $f_\mathrm{H}$.
    \item The GDH features traced by the spiral arms are not due to the star formation feedback, such as \HII\ regions, because the star formation feedback can produce an effect on its GDH only in a smaller-scale, and it has little effect on our GDH.
    \item These results provide information about what happens in the spiral arms. The interstellar shock waves, both supersonic and subsonic, will affect the structural evolution of the ISM. The large-scale strong shock will produce a peak density shift of an LN component in the GDH, and the small-scale weak shock or subsonic pressure modification will result in a wider density dispersion of the low-density LN component. This means that the GDH is a powerful tool to investigate the density evolution of the ISM, although we need finer simulations.

\end{enumerate}
This work suggests that the GDH analysis is useful to address the ISM response to large-scale dynamics induced by galactic potential. We can confirm the relation between the GDH parameters and spiral arms based on this GDH analysis using high-fidelity CO images of nearby galaxies observed with the Atacama Large Millimeter/submillimeter Array \citep[e.g.][]{LeroyETAL2021,KodaETAL2023,MuraokaETAL2023}. This should open the door to a comprehensive picture of the evolution of the ISM in a galaxy.

\section*{Acknowledgements}
We thank the reviewer for the careful reading and helpful comments that contributed to improve this article.  We are grateful to Drs. H. Imai, M. Kobayashi, and M. Kohno or the useful discussions. The 45-m radio telescope is operated by the Nobeyama Radio Observatory, a branch of the National Astronomical Observatory of Japan. This publication makes use of data from FUGIN, the FOREST Unbiased Galactic plane imaging survey with the Nobeyama 45-m telescope, a legacy project in the Nobeyama 45-m radio telescope. This research used \textsc{astropy},\footnote{\url{http://www.astropy.org}} a community-developed core Python package for Astronomy \citep{astropy_2013, astropy_2018}, \textsc{matplotlib}, a Python package for visualisation \citep{Hunter2007}, \textsc{numpy}, a Python package for scientific computing \citep{HarrisETAL2020}, and Overleaf, a collaborative tool. We would like to thank Editage (http://www.
editage.com) for English language editing.

\section*{Data Availability}
The data underlying this article are available in the article and the public data release of the FUGIN survey \citep{UmemotoETAL2017}.
 



\bibliographystyle{mnras}
\bibliography{reference} 

\appendix

\bsp	
\label{lastpage}
\end{document}